\documentclass[a4paper,11pt]{article}
\pdfoutput=1 

\usepackage{jcappub} 
\usepackage[toc,page]{appendix}
\usepackage{subcaption}
\usepackage{amsmath}
\usepackage{graphicx}                
\usepackage{multirow}
\usepackage[T1]{fontenc} 
\usepackage[table]{xcolor}
\usepackage{float}
\usepackage{orcidlink}
\usepackage{bm}

\newcommand{\DNCMBinst}{\Delta N_\text{CMB, inst rh}}
\newcommand{\DNrh}{\Delta N_\text{rh}}
\newcommand{\kCMB}{k_\text{CMB}}
\newcommand{\phiCMB}{\phi_\text{CMB}}
\newcommand{\xCMB}{x_\text{CMB}}
\newcommand{\xend}{x_\text{end}}
\newcommand{\DNCMB}{\Delta N_\text{CMB}}

\hypersetup{pdftitle={Beyond monomial alpha-attractors}
}

\title{\boldmath 
Beyond monomial $\alpha$-attractors}

\author[a,b]{Laura Iacconi\,$^{\text{\orcidlink{0000-0002-1152-3056}}}$}
\affiliation[a]{Astronomy Unit, Queen Mary University of London, \\ Mile End Road, London, E1 4NS, U.K.}
\affiliation[b]{Institute of Cosmology \& Gravitation, University of Portsmouth,\\ Burnaby Road, Portsmouth, PO1 3FX, U.K.}

\emailAdd{l.iacconi@qmul.ac.uk}

\abstract{
Recent small-scale CMB data show a preference for larger scalar spectral index, $n_s$, when combined with DESI data. 
Monomial $\alpha$-attractor T-models can be reconciled with the new observations if the power of the hyperbolic tangent function is $p\geq 6$, where $p$ is even. 
The supergravity construction of monomial T-models with $p>2$ relies on the assumption that lower powers remain negligible over the whole field range explored during inflation and reheating. 
What would be the consequences of going beyond the monomial formulation? 
As a first step in this direction, we consider the case of a binomial potential, given by the sum of quadratic and quartic terms. 
When the quartic coefficient, $c$, is $0< c+1/2 \ll 1$, we find that the new model displays non-universal behavior for $n_s$, leading to values as large as $0.965$ and with an $\alpha$-dependence that is qualitatively different from that of monomial potentials. 
By solving the background dynamics during the first few e-folds of perturbative reheating we show that the quartic term might dominate before the quadratic one eventually takes over as the inflaton oscillations decrease in amplitude.
This leads to a time-dependent equation of state, with $\bar w\sim 1/3$ initially before ultimately $\bar w\to 0$. 
Obtaining a quartic-dominated reheating stage lasting $\sim4$ e-folds requires a substantial hierarchy between the quartic and quadratic terms, $c\sim 10^5$ for $\alpha\gtrsim 0.1$. 
Our study highlights that models beyond the monomial form can lead to non-trivial deviations of $n_s$ from the predictions of monomial potentials. 
Furthermore our results for $\bar w$ during reheating call into question the use of monomial models with large $p$; assuming that $\bar w$ is uniquely determined by the $p$-\textit{th} power relies on substantial fine-tuning of the underlying supergravity potential. 
}

\begin{document}
	\maketitle
	\flushbottom
	
\section{Introduction}
\label{sec: introduction}
Cosmological inflation is the leading paradigm to describe the very early universe~\cite{Liddle:2000cg}. 
Initially proposed as a solution to the horizon problem, quantum vacuum fluctuations generated during inflation also explain the origin of primordial perturbations observed in the Cosmic Microwave Background (CMB). 

The accelerated expansion of the background characterising inflation can be realised within the simplest models through the dynamics of a single, canonical scalar field, the inflaton, slowly rolling down its potential, $V(\phi)$.
The landscape of single-field models of inflation is highly populated~\cite{Martin:2013tda, Martin:2024qnn}. 
Amongst these, cosmological $\alpha$-attractors stand out as particularly compelling potentials, see Ref.~\cite{Kallosh:2015zsa} and references therein.
In this work our starting point is a particular subclass of $\alpha$-attractors, T-models~\cite{Kallosh:2013hoa}
\begin{equation}
    \label{eq: potential monomial T-models}
    V(\phi)
    =
    V_0 
    \tanh^p\left(\frac{\phi}{\sqrt{6\alpha}}\right) \;, 
\end{equation}
where $p$ is an even integer $p\geq 2$ and we have set the reduced Planck mass, $M_\text{Pl}\equiv(8\pi G_N)^{-1/2}$, to unity.
To explain the appealing qualities of cosmological $\alpha$-attractors, we will therefore use T-models as an example. 

Cosmological $\alpha$-attractors can be formulated within supergravity theories in terms of a complex scalar field, $Z\equiv \tanh\left({\phi}/{\sqrt{6\alpha}}\right)\,e^{i\theta}$, belonging to the
Poincaré hyperbolic disk~\cite{Carrasco:2015uma, Kallosh:2015zsa}. 
By assuming that the potential $V(Z,\, \bar Z)$ is invariant under a phase shift $Z\to e^{i\varphi}Z$, only powers of $Z\bar Z$ can appear, i.e. even powers of $\tanh( {\phi}/{\sqrt{6\alpha}})$.
In terms of the canonical inflaton field, $\phi$, the (simplest version of the) supergravity potential reads 
\begin{equation}
\label{eq: potential full series}
  V(\phi)
  =
  c_2\tanh^2\left( \frac{\phi}{\sqrt{6\alpha}}\right)
  +
  c_4\tanh^4\left( \frac{\phi}{\sqrt{6\alpha}}\right)
  +
  c_6\tanh^6\left( \frac{\phi}{\sqrt{6\alpha}}\right)
  +
  \mathcal{O}\left[\tanh^8\left( \frac{\phi}{\sqrt{6\alpha}}\right)\right] \;. 
\end{equation}
By assuming that the $p$-\textit{th} power dominates over the whole range explored by $\phi$ during inflation \textit{and} reheating, Eq.~\eqref{eq: potential full series} yields the monomial T-model in Eq.~\eqref{eq: potential monomial T-models}. 

The predictions of cosmological $\alpha$-attractors for large-scale observables are \textit{universal}, i.e. largely insensitive to the potential parameters. 
By labeling the number of e-folds elapsed between the horizon crossing of the CMB pivot scale to the end of inflation as $\DNCMB$, in a large-$\DNCMB$ expansion the scalar spectral tilt and the tensor-to-scalar ratio are given by~\cite{Kallosh:2013hoa, Kallosh:2013yoa} 
\begin{equation}
    \label{eq: universal predictions}
    n_s = 1-\frac{2}{\DNCMB} + \mathcal{O}(\DNCMB^{-2}) \;, \quad r = \frac{12\alpha}{\DNCMB^2} + \mathcal{O}(\DNCMB^{-3}) \;. 
\end{equation}
While of course $\DNCMB$ is determined by the inflationary potential (and details of the reheating phase), the dependence on $V(\phi)$ is only logarithmic, see Eq.~\eqref{eq:DNCMB-general}.
Therefore, $n_s$ is largely independent of the potential parameters $\alpha$ and $p$, while $\alpha$ sets the magnitude of $r$, the hallmark of $\alpha$-attractor models.  

For $50\lesssim\DNCMB \lesssim 60$ and $\alpha \lesssim \mathcal{O}(1)$, the predictions~\eqref{eq: universal predictions} sit comfortably within the bounds from CMB observations by \textit{Planck}~\cite{Planck:2018jri} and BICEP-Keck 2018~\cite{BICEP:2021xfz}. 
Nevertheless, recent smaller-scale CMB data from ACT~\cite{AtacamaCosmologyTelescope:2025blo} and SPT-3G~\cite{SPT-3G:2025bzu} show a preference for larger values of $n_s$ when used in combination with DESI BAO data~\cite{DESI:2025zgx}. 
This shift is driven by a tension between BAO parameters as determined from CMB and DESI data~\cite{SPT-3G:2025bzu, Ferreira:2025lrd}, and it needs therefore to be treated with caution.  
However, if taken at face value the preference for larger $n_s$ puts previously preferred inflationary models, such as $\alpha$-attractors with standard reheating ($0\leq \bar w \leq 1/3$, where $\bar w$ is the equation of state parameter during reheating), in tension with the new measurements. 
It has been shown in Refs.~\cite{Drees:2025ngb, Haque:2025uri, Haque:2025uga, German:2025ide, Ellis:2025zrf, Iacconi:2025odq} that the discrepancy between new measurements and monomial T-model potentials~\eqref{eq: potential monomial T-models} can be alleviated if $p\geq6$. 
In this case, T-models can self-consistently produce an extended reheating stage with stiff equation of state ($ \bar w > 1/3$), which allows values of $n_s$ closer to unity~\cite{Iacconi:2025odq}. 

From a model-building perspective, one might question whether monomial T-models with large $p$ are theoretically motivated. 
Assuming Eq.~\eqref{eq: potential monomial T-models} with $p\geq 4$ requires the lower powers contributing to Eq.~\eqref{eq: potential full series} to remain negligible at all times, even during reheating when $|\phi|/\sqrt{6\alpha} \ll 1$ would naturally enhance their importance. 
What are the consequences of working directly with Eq.~\eqref{eq: potential full series}?
As a first step in this direction, in this work we consider the binomial potential
\begin{equation}
\label{eq:potential with two terms}
  V(\phi)
  = V_0\left[
  \tanh^2\left(\frac{\phi}{\sqrt{6\alpha}}\right)
  + c\,\tanh^4\left(\frac{\phi}{\sqrt{6\alpha}}\right)
  \right] \;, 
\end{equation}
where we have set $V_0\equiv c_2$ and $c\equiv c_4/c_2$.
Note that in the limit $c\to 0$ ($c\to \infty$) one recovers the $p=2$ ($p=4$) monomial potential~\eqref{eq: potential monomial T-models}. 
We show that novel non-universal behavior of $n_s$ might be produced from Eq.~\eqref{eq:potential with two terms}, even if the building blocks of the potential are two monomial T-models. 
Furthermore, we show that Eq.~\eqref{eq:potential with two terms} might yield a time-dependent equation of state during reheating, with an initial phase dominated by the quartic term before the quadratic one takes over. 
A very large hierarchy between $c_4$ and $c_2$ is required to have even just a few e-folds dominated by the quartic term.  

\medskip
\textit{Content:} 
In Sec.~\ref{sec: binomial potential during inflation} we obtain the large-scale predictions for the potential in Eq.~\eqref{eq:potential with two terms}. 
First, we review how $n_s$ and $r$ are computed in Sec.~\ref{sec: background and ns and r review}, and then provide results for instantaneous reheating for $c>0$ and $-1/2\leq c<0$ in Secs.~\ref{sec: binomial potential c>0} and~\ref{sec: binomial potential c<0} respectively. 
In Sec.~\ref{sec: reheating} we study the post-inflationary dynamics, and determine the impact of $c$ on the equation of state parameter during the first few e-folds of perturbative reheating. 
In Sec.~\ref{sec:evolvingw} we numerically compute $\bar w$ for three models with sizeable $c$, explicitly showing that the equation of state is time-dependent due to the interplay between the quartic and quadratic terms in Eq.~\eqref{eq:potential with two terms}. 
We then perform a systematic parameter-space exploration to establish how large $c$ needs to be in order to realise a transient reheating stage dominated by the quartic term in Sec.~\ref{sec: fine tuning for w>=1/3}.
We present our conclusions in Sec.~\ref{sec:conclusions}. 
In order to establish at what order in the large-$\DNCMB$ expansion the parameters $c$ and $\alpha$ enter the $n_s$ results, we compute analytical predictions for $n_s$ in App.~\ref{app: analytical predictions for ns}. 
The results of App.~\ref{app: analytical predictions for ns} complement those of Sec.~\ref{sec: binomial potential during inflation}. 

\medskip
\textit{Conventions:} 
Throughout this work, we consider a spatially-flat Friedmann--Lema\^{i}tre--Robertson--Walker universe, with line element $\text{d}s^2=-\text{d}t^2+a^2(t)\delta_{ij}\text{d}x^i\text{d}x^j$, where $t$ denotes cosmic time and $a(t)$ is the scale factor. The Hubble rate is defined as $H\equiv {\dot a}/{a}$,  where a derivative with respect to cosmic time is denoted by $\dot f \equiv {\mathrm{d}f}/{\mathrm{d}t}$. The number of e-folds of expansion is defined as $N\equiv \int \mathrm{d}t\,H(t)$ and $f'\equiv {\mathrm{d}f}/{\mathrm{d}N}$. 

\section{Inflationary dynamics and large-scale observables}
\label{sec: binomial potential during inflation}

In this section we focus on the inflationary dynamics of the binomial potential in Eq.~\eqref{eq:potential with two terms}. 
First, we find a condition on $c$ that ensures monotonicity of $V(\phi)$, such that inflation can be successfully realised for arbitrary large initial values of $\phi$. 
By differentiating Eq.~\eqref{eq:potential with two terms} and neglecting positive coefficients we obtain
\begin{equation}
    V_{\phi} \equiv \frac{\mathrm{d}V(\phi)}{\mathrm{d}\phi}
    \propto
    \left[
    1 - 2c + 
    (1+2c) \cosh\left(\frac{2\,\phi}{\sqrt{6\alpha}}\right)
    \right]
    \operatorname{sech}^{4}\left(\frac{\phi}{\sqrt{6\alpha}}\right)
    \,
    \tanh\left(\frac{\phi}{\sqrt{6\alpha}}\right) \;. 
  \label{eq:V_phi binomial}
\end{equation}
The potential is symmetric under $\phi\to -\phi$, so it is sufficient to consider $\phi>0$. 
For $\phi\to 0^+$ and $\phi\to \infty$ the potential flattens $(V_\phi\to 0)$, and monotonicity with $V_\phi>0$ at intermediate values is ensured if
\footnote{When $c<-1/2$, the potential develops a maximum at $\phi = \sqrt{{3\alpha}}/\sqrt{2}\, \operatorname{arcosh}\left[(2c-1)/(2c+1)\right]$.}
$c\geq -1/2$. 
The absolute minimum lies at $\phi=0$, where reheating may occur.  

After reviewing briefly how numerical predictions for the scalar spectral tilt, $n_s-1$, and tensor-to-scalar ratio, $r$, are obtained in Sec.~\ref{sec: background and ns and r review}, we will provide $n_s$ and $r$ computed for instantaneous reheating for models with $c>0$ and $-1/2\leq c<0$ in Sec.~\ref{sec: binomial potential c>0} and~\ref{sec: binomial potential c<0} respectively.

\subsection{Background evolution and CMB observables}
\label{sec: background and ns and r review}

For each inflationary model~\eqref{eq:potential with two terms} in the $(\alpha,\, c)$ parameter space we numerically solve the background evolution, determined by the equations 
\begin{equation}
\label{eq: background evolution}
    \ddot \phi + 3 H\dot \phi + V_\phi =0 
    \quad
    \text{and}
    \quad 
    \dot H + \frac{1}{2} \dot \phi^2 =0 \;, 
\end{equation}
where $\phi$ and $H$ obey the Friedmann constraint 
\begin{equation}
\label{eq: Friedmann constraint}
    H^2 = \frac{1}{3} \left[\frac{1}{2} \dot \phi^2 + V(\phi) \right] \;. 
\end{equation}
We solve Eqs.~\eqref{eq: background evolution} in terms of e-folding number, $\mathrm{d} N \equiv H \mathrm{d} t$.  
The initial field value, $\phi_\text{in}$, is chosen such that inflation lasts for at least 90 e-folds, ensuring that the background has reached the attractor solution by the time the CMB scale exited the horizon.
The initial velocity is set to zero, $\phi'_\text{ in}=0$ (a prime denotes a derivative with respect to $N$), and the initial Hubble rate is fixed by the Friedmann constraint~\eqref{eq: Friedmann constraint}. 
The differential solver is interrupted when $\epsilon_H\equiv -H'/H=1$, signaling the end of inflation. 

We obtain large-scale predictions for the scalar spectral tilt, $n_s-1$, and the tensor-to-scalar ratio, $r$, at second-order in the slow-roll expansion
\footnote{First-order slow-roll predictions for $n_s$ are not sufficiently precise for comparing theoretical predictions against current CMB measurements, see e.g. Ref.~\cite{Martin:2013tda}. 
While we employ expressions in terms of potential slow-roll parameters, see Eq.~\eqref{eq:slowroll-parameters}, we have checked that these are indistinguishable from those obtained with Hubble slow-roll parameters, see Eqs.~(B.3)-(B.6) in Ref.~\cite{Iacconi:2023mnw}.}
\begin{align}
\label{eq:nssecond}
  n_s-1
  &= 
  -6\epsilon_V+2\eta_V
  +4\left(\frac{11}{3}-\frac{3}{2}C\right)\epsilon_V^2
  -2(7-2C)\epsilon_V\eta_V 
  +\frac{2}{3}\eta_V^2
  +\frac12\left(\frac{13}{3}-C\right)\xi_V^2 + \mathcal{O}(\epsilon^3) \;, 
  \\
  \label{eq:r second}
  r
  &=
  16\,\epsilon_V 
    \left[ 1 + \left(\frac{C}{2} - \frac{13}{6}\right)\left(2\epsilon_V - \eta_V\right) \right] + \mathcal{O}(\epsilon^3) \;, 
\end{align}
where $C=4\log 2+4\gamma_{\rm E}-5$ and $\gamma_{\rm E}$ is Euler-Mascheroni's constant. 
Here the potential slow-roll parameters are defined as 
\begin{equation}
  \epsilon_V \equiv \frac{1}{2}
  \,
  \left(\frac{V_{\phi}}{V}\right)^2
  \;, 
  \qquad
  \eta_V \equiv \frac{V_{\phi\phi}}{V} 
  \;, 
  \qquad
  \xi_V^2 \equiv \frac{V_{\phi}V_{\phi\phi\phi}}{V^2} \;. 
  \label{eq:slowroll-parameters}
\end{equation}
Numerical results for $n_s$ and $r$ are obtained by substituting the numerical background solutions in Eqs.~\eqref{eq:nssecond} and~\eqref{eq:r second}, and evaluating them when the CMB pivot scale, $\kCMB=0.05\,\text{Mpc}^{-1}$, exited the horizon. 
The number of e-folds elapsed between the horizon crossing of $\kCMB$ and the end of inflation is~\cite{Planck:2018jri}
\footnote{Here we have set the effective number of relativistic degrees of freedom at the end of reheating to $10^3$ and $H_0$ to the central value as measured by \textit{Planck}, $H_0 = (67.4\pm 0.5) \, \text{km} \, \text{s}^{-1}\, \text{Mpc}^{-1}$~\cite{Planck:2018vyg}.}
\begin{equation}
\label{eq:DNCMB-general}
\begin{split}
  \DNCMB 
  &\equiv 
  N_\text{end}- N_\text{CMB} \\
  & =
  \underbrace{60.7
  +
  \frac{1}{4}\ln\!\left(\frac{V_{\rm CMB}^2}{\rho_{\rm end}\, M_\text{Pl}^4}\right)}_{\equiv \DNCMBinst}
  -\frac{1-3\bar w}{4} \DNrh  \;,     
\end{split}
\end{equation}
where $V_{\rm CMB}$ is the potential evaluated at CMB crossing, $\rho_{\rm end}$ is the energy density at the end of inflation, $\bar w$ is the effective reheating equation of state and $\DNrh$ is the duration of reheating. 
To obtain $\DNCMBinst$, i.e. the value of $\DNCMB$ for instant reheating, we iteratively solve Eq.~\eqref{eq:DNCMB-general} with $\DNrh=0$ and for
values of $V_0$ compatible with CMB observations. 
In Secs.~\ref{sec: binomial potential c>0} and~\ref{sec: binomial potential c<0} we present results for $n_s$ and $r$ computed for instant reheating, and discuss reheating separately in Sec.~\ref{sec: reheating}. 

\subsection[\texorpdfstring{Results for $c>0$}{Results for c>0}]{Results for $\bm{c>0}$}
\label{sec: binomial potential c>0}

In this section, we consider the area of parameter space defined by $10^{-4}\leq \alpha \leq 10$ and $10^{-2}\leq c\leq 10^2$, with logarithmic priors for both potential parameters~\cite{Iacconi:2023mnw}.

\begin{figure}
  \centering
  \includegraphics[width=0.7\textwidth]{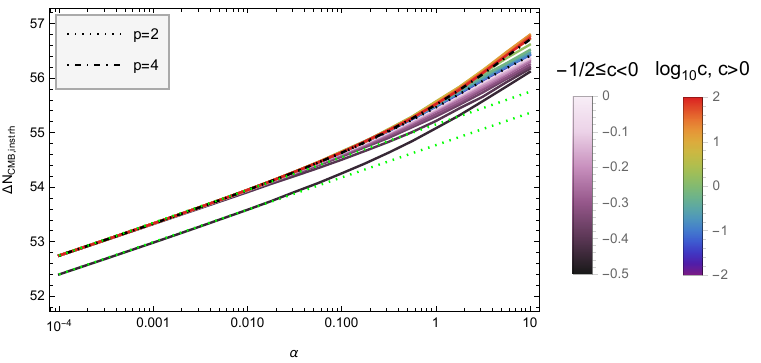}
  \caption{Values of $\DNCMBinst$, see Eq.~\eqref{eq:DNCMB-general}, shown against $\alpha$ for binomial models~\eqref{eq:potential with two terms} with $c>0$ (rainbow-coloured lines) and $-1/2\leq c<0$ (purple lines).
  Black lines show results for monomial models~\eqref{eq: potential monomial T-models} with $p=2$ (dotted) and $p=4$ (dot-dashed).
  Note the dark-purple line, below the black ones, corresponds to $c=-1/2$.  
  The dotted, green lines correspond to the functions $55.14+ 0.6 \log_{10}\alpha$ and $54.77 + 0.6 \log_{10}\alpha$, which encode the linear dependence of $\DNCMBinst$ on $\log_{10}\alpha$ when $\alpha\lesssim 0.1$.}
  \label{fig: DN_CMB inst rh binomial models}
\end{figure}

Values of $\DNCMBinst$ are presented in Fig.~\ref{fig: DN_CMB inst rh binomial models}, together with results for models with $-1/2\leq c<0$. 
As for monomial T-models, we observe that $\DNCMBinst$ displays a logarithmic dependence on $\alpha$~\cite{Iacconi:2023mnw}. 
For $\alpha\lesssim 0.1$, $\DNCMBinst$ does not depend on the additional potential parameter ($p$ for monomial T-models and $c$ for the binomial potential) and scales linearly with $\log_{10}\alpha$, see the dotted-green line. 
For larger $\alpha$ the extra parameter becomes relevant and it induces a running on the dependence of $\DNCMBinst$ on $\log_{10}\alpha$. 
This can be seen from the deviation of the coloured and black lines from the dotted-green one. 
Note that even for the largest $\alpha$ we consider ($\alpha=10$) the variation in $\DNCMBinst$ due to the additional potential parameter is at most one e-fold. 
For potentials with $c\sim 0.01$ ($c\sim 100$), $\DNCMBinst$ approaches $p=2$ ($p=4$) results.  
Nevertheless, for intermediate values of $c$ the binomial potential~\eqref{eq:potential with two terms} does not simply monotonically interpolate between $p=2$ and $p=4$ models, as we will describe in details below for $n_s$. 

\begin{figure}
  \centering
  \includegraphics[width=0.45\textwidth]{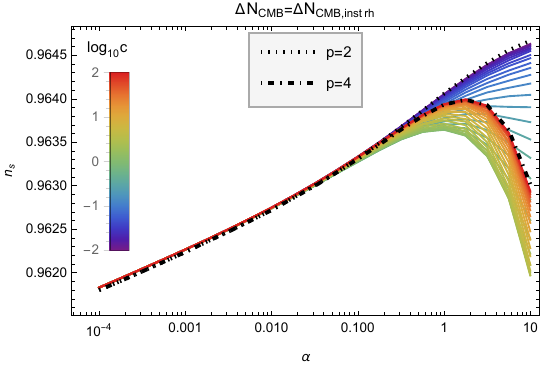}
  \includegraphics[width=0.46\textwidth]{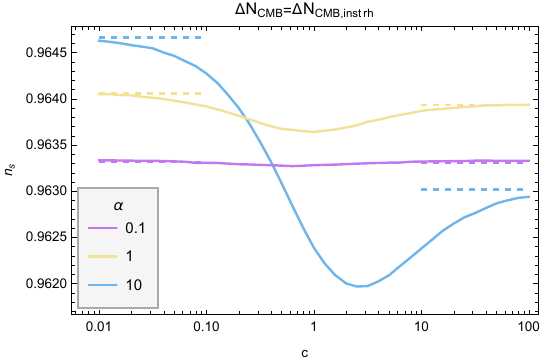}
  \caption{\textit{Left panel:} Scalar spectral tilt, $n_s$, shown as a function of $\alpha$ for binomial models~\eqref{eq:potential with two terms} with $c>0$. 
  See the bar legend for values of $\log_{10}c$. 
  The black lines display $n_s$ values computed for monomial T-models~\eqref{eq: potential monomial T-models} with $p=2$ (dotted) and $p=4$ (dot-dashed).
  \textit{Right panel:}
  $n_s$ shown as a function of $c$ for three values of $\alpha$, see the legend. 
  For each $\alpha$, dashed lines mark the value of $n_s$ computed for monomial models with $p=2$ (displayed for small $c$) and $p=4$ (displayed for large $c$). 
  In both panels, $n_s$ has been computed numerically from Eq.~\eqref{eq:nssecond} under the assumption of instantaneous reheating, $\DNCMB=\DNCMBinst$.}
  \label{fig:ns-alpha}
\end{figure}

Numerical results for $n_s$ computed by assuming instantaneous reheating are displayed in Fig.~\ref{fig:ns-alpha}. 
As for $\DNCMBinst$, we observe in the left panel that for $\alpha\lesssim0.1$ the value of $c$ has negligible impact, while for larger $\alpha$ $n_s$ approaches T-model results with $p=2$ ($p=4$) when $c=0.01$ ($c=100$). 
While for asymptotically small and large $c$ we recover the appropriate monomial predictions, models with intermediate $c$ values do not linearly interpolate between the two regimes, see the right panel of Fig.~\ref{fig:ns-alpha}. 
For $\alpha=10$ we observe that for increasing $c$ at first $n_s$ decreases from the $p=2$ value and then increases and approaches the $p=4$ prediction from below. 
The larger the value of $\alpha$, the more pronounced the dependence of $n_s$ on $c$ is.    

While for large $\alpha$ the $c$-dependence of $n_s$ is non-trivial, the departure from monomial models with $p=2$ and $p=4$ remains at most of $\mathcal{O}(10^{-3})$ over the range of $\alpha$ and $c$ displayed in Fig.~\ref{fig:ns-alpha}.
The largest deviation is produced for models with $\alpha=10$ and $c\sim3$, for which the relative difference with respect to $n_s$ produced with $p=4$ is $\sim11\%$.
For fixed $c$, the dependence of $n_s$ on $\alpha$ is controlled by the magnitude of $c$, with $n_s$ increasing monotonically for increasing $\alpha$ when $c\ll1$ and displaying a maximum when $c$ is sizeable.
Nevertheless, such $\alpha$-dependence is qualitatively the same as for monomial models with $p=2$ and $p=4$, and for $c>0$ $n_s$ is always smaller than the $p=2$ results. 
The fact that $n_s$ predictions are not substantially sensitive to the potential parameters $c$ and $\alpha$ suggests that $\alpha$-attractor binomial models~\eqref{eq:potential with two terms} with $c>0$ still display universal behavior~\cite{Kallosh:2013hoa,Kallosh:2013yoa}. 
We explicitly show that $n_s$ predictions are universal at leading order in the large-$\DNCMB$ expansion in App.~\ref{app: analytical predictions for ns with c>0}
\footnote{Note that the results presented in Fig.~\ref{fig:ns-alpha} and Fig.~\ref{fig: ns analytical c>0} are different as in the former $n_s$ is evaluated at $\DNCMB=\DNCMBinst$ (which itself depends on $\alpha$ and $c$), while in the latter it is computed for fixed $\DNCMB=55$.}. 
We report here the resulting expression,  Eq.~\eqref{eq: ns large DNCMB c>0}, 
\begin{multline}
  n_s = 1-\frac{2}{\Delta N_{\rm CMB}}
  +\frac{1}{{\Delta N_{\rm CMB}}^2 4(1+2c)^2}
  \Bigg[
  3 \alpha (1+c) (1+2c)
  \exp{\left(2 \xend\right)}
  +24\alpha c\,x_{\rm end}
    \\
  -6\alpha\left(1+c(4c+\ln 64)\right)
  +12\alpha c \ln\left(
  \frac{3\alpha(1+c)}{(1+2c)\Delta N_{\rm CMB}}
  \right)
  \Bigg]
  +\mathcal{O}\left({\Delta N_{\rm CMB}}^{-3}\right) \;. 
\end{multline}
The leading term reproduces universal $\alpha$-attractor predictions, see Eq.~\eqref{eq: universal predictions}, while the dependence on $c$ and $\alpha$ enters at $\mathcal{O}(\DNCMB^{-2})$. 

\begin{figure}
  \centering
  \includegraphics[width=0.45\textwidth]{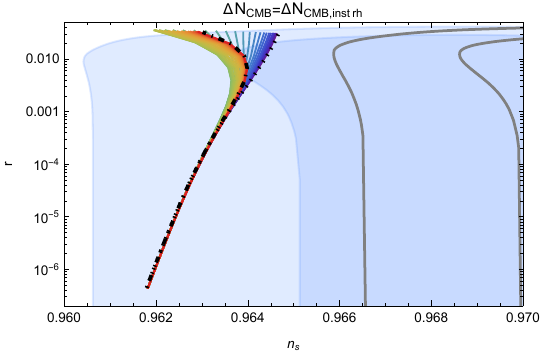}
  \centering
  \includegraphics[width=0.45\textwidth]{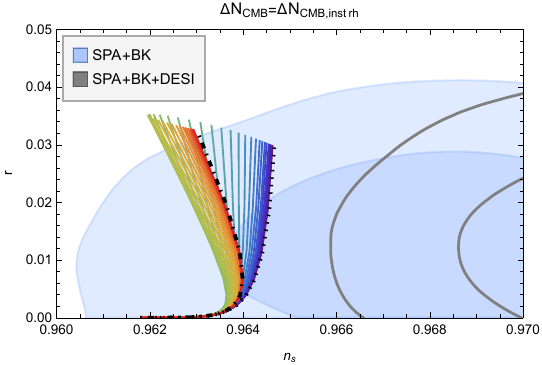}
  \caption{Numerical predictions in the $(n_s, \, r)$ plane for binomial models~\eqref{eq:potential with two terms} with $c>0$. 
  Values of $n_s$ and $r$ have been computed numerically from Eqs.~\eqref{eq:nssecond} and~\eqref{eq:r second} under the assumption of instantaneous reheating, $\DNCMB=\DNCMBinst$. 
  See Fig.~\ref{fig:ns-alpha} for the colour legend identifying different $c$ values.
  Moving from bottom (smaller $r$) to top (larger $r$) along each fixed-$c$ coloured line corresponds to increasing $\alpha$ from $10^{-4}$ to $10$.
  The black lines show $n_s$ and $r$ values computed for monomial T-models~\eqref{eq: potential monomial T-models} with $p=2$ (dotted) and $p=4$ (dot-dashed).
  Results for $n_s$ and $r$ are displayed together with the SPA+BK and SPA+BK+DESI $68\%$ and $95\%$ C.L. contours~\cite{Balkenhol:2025wms}.
  In the left (right) panel $r$ is shown on a logarithmic (linear) scale.}
  \label{fig:ns-r binomial models with c>0}
\end{figure}

Predictions in the $(n_s, r)$ plane are displayed in Fig.~\ref{fig:ns-r binomial models with c>0}. 
We find that $r$ is mainly controlled by $\alpha$, as in ordinary $\alpha$-attractor models, with models with $c=0.01$ ($c=100$) approaching the $p=2$ ($p=4$) predictions.
As for $n_s$, intermediate values of $c$ interpolate non-monotonically between $p=2$ and $p=4$ predictions. 
This can be seen by considering the endpoints of the coloured lines in the right panel, which correspond to models with $\alpha=10$ and different values of $c$. 

In Fig.~\ref{fig:ns-r binomial models with c>0}, $n_s$ and $r$ predictions are compared against the latest constraints from large-scale CMB observations and BAO measurements. 
SPA+BK contours are obtained from combining \textit{Planck}, SPT, ACT and BICEP/Keck 2018 data; adding also DESI BAO data yields the SPA+BK+DESI contours. 
The MCMC chains used to produce the aforementioned contours are supplied in Ref.~\cite{Balkenhol:2025wms}. 
Note that once BAO DESI data are added, CMB datasets show a preference for larger $n_s$ values. 
This is due to the correlation between $n_s$ and BAO parameters, and the tension between BAO parameters as determined from CMB and DESI data~\cite{Ferreira:2025lrd}.  
For the range of $\alpha$ considered, binomial models with $c>0$ and $\DNCMB=\DNCMBinst$ are mostly compatible with SPA+BK at least at $95\%$ C.L., although some large-$\alpha$ models leave the contours, see the right panel of Fig.~\ref{fig:ns-r binomial models with c>0}. 
By contrast, the addition of DESI BAO data and the consequent shift in the preferred $n_s$ to larger values place these instantaneous-reheating predictions in tension with the SPA+BK+DESI data at more than $95\%$ C.L.. 
Considering that $n_s$ predictions for binomial models with $c>0$ are below those of monomial potentials with $p=2$ (see the left panel of Fig.~\ref{fig:ns-alpha}), the tension with SPA+BK+DESI data found here is consistent with the analysis of monomial models done in Ref.~\cite{Iacconi:2025odq}. 

\subsection[\texorpdfstring{Results for $-1/2\leq c<0$}{Results for -1/2<=c<0}]{Results for $\bm{-1/2\leq c<0}$}
\label{sec: binomial potential c<0}

In this section we consider the area of parameter space defined by $10^{-4}\leq \alpha \leq 10$ and $-1/2\leq c<0$. 
Unlike in Sec.~\ref{sec: binomial potential c>0}, we now employ a linear prior for $c$. 

Values of $\DNCMBinst$ are shown in Fig.~\ref{fig: DN_CMB inst rh binomial models}, together with those computed for models with $c>0$. 
For $c\neq -1/2$ and small $\alpha$, $\DNCMBinst$ scales linearly with $\log_{10}\alpha$ and shows no dependence on $c$. 
For larger $\alpha$, $c$ induces a running in the slope of $\DNCMBinst$, making it slightly steeper. 
Models with $|c|\ll 1$ closely track the values computed for T-models with $p=2$. 
For $c=-1/2$ and small $\alpha$, $\DNCMBinst$ scales linearly with $\log_{10}\alpha$, with the same slope as for models with $c\neq -1/2$ but with a smaller intercept (see the definitions of the green-dashed lines in the caption of Fig.~\ref{fig: DN_CMB inst rh binomial models}). 
Models with $c\to -1/2$ interpolate between the $p=2$ and $c=-1/2$ predictions, with the $\alpha$ value for which the binomial model deviates from the $p=2$ line being smaller as $c$ approaches $-1/2$. 
For example, when $c=-0.45$ this is realised for $\alpha=\mathcal{O}(10^{-2})$ and when $c=-0.495$ for $\alpha=\mathcal{O}(10^{-3})$. 

\begin{figure}
  \centering
  \includegraphics[width=0.42\textwidth]{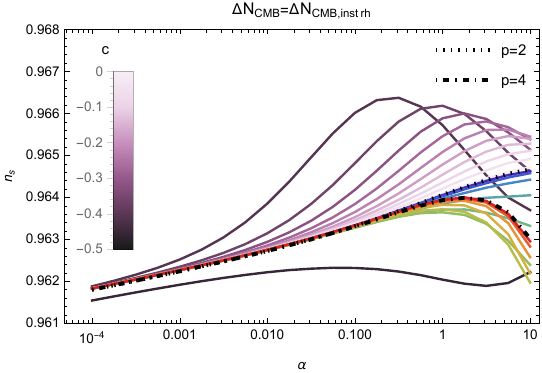}
  \centering
  \includegraphics[width=0.49\textwidth]{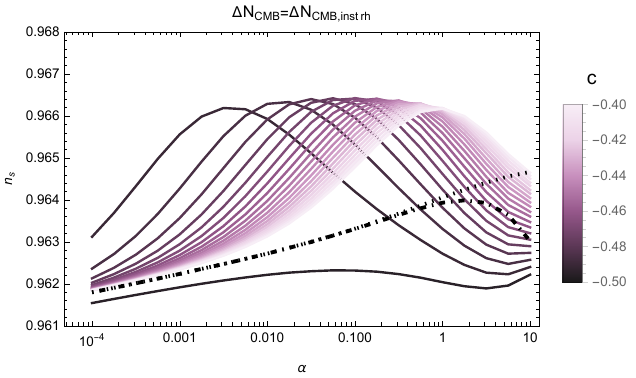}
  \caption{\textit{Left panel:} Scalar spectral index, $n_s$, shown as a function of $\alpha$ for binomial models~\eqref{eq:potential with two terms} with $c>0$ (rainbow-colour lines) and $-1/2\leq c<0$ (purple lines). 
  See Fig.~\ref{fig: DN_CMB inst rh binomial models} for the legend identifying models with $c>0$. 
  The black lines correspond to $n_s$ values computed for monomial T-models~\eqref{eq: potential monomial T-models} with $p=2$ (dotted) and $p=4$ (dot-dashed). 
  Note the dark-purple line at the bottom of the plot represents models with $c= -1/2$. 
  \textit{Right panel:} Values of $n_s$ computed for binomial models with $-1/2\leq c \leq -0.4$, see the bar legend. 
  See the left panel for the legend identifying the T-models predictions. 
  Note that the purple lines correspond to different $c$ values with respect to the left panel. 
  In both panels, $n_s$ is computed by assuming instantaneous reheating.}
  \label{fig: ns binomial with c<0}
\end{figure}

In Fig.~\ref{fig: ns binomial with c<0} we show values of the scalar spectral index, $n_s$. 
In the left panel we populate the range $-1/2\leq c<0$ with 11 models. 
We observe that for $|c|\to0$ values of $n_s$ closely track the $p=2$ results, with $n_s$ monotonically increasing for increasing $\alpha$.
Predictions for models with $c=-1/2$ display quite a different $\alpha$-dependence with respect to monomial models with $p=2$ and $p=4$. 
$n_s$ is rigidly shifted downwards when $\alpha=\mathcal{O}(10^{-4})$.
While for monomial models $n_s$ increases monotonically for intermediate $\alpha$ values and either remains monotonic ($p=2$) or starts decreasing when $\alpha\approx 2$ ($p=4$), when $c=-1/2$ we observe a shallower profile, with an earlier local maximum at $\alpha\approx 0.1$, a local minimum at $\alpha\approx 3$, followed by a last increase. 

As $c$ decreases towards $-1/2$, $n_s$ changes its monotonic $\alpha$-dependence, developing first a local maximum located at smaller and smaller $\alpha$, and a minimum as well when $c\to -1/2$. 
See the right panel of Fig.~\ref{fig: ns binomial with c<0} for a zoom-in over the range $-0.4\leq c\leq -0.5$. 
For $c\to -1/2$ the $n_s$ predictions at large $\alpha$ asymptotically approach those of $c=-1/2$ models, while for smaller $\alpha$ values there is a sizeable difference between the two. 
For example, for $\alpha=3 \times 10^{-3}$ a model with $c=-0.495$ leads to $n_s\approx 0.966$, with a difference of $4\times 10^{-3}$ with respect to the $c=-1/2$ result, $n_s\approx0.962$.

\begin{figure}
  \centering
  \includegraphics[width=0.55\textwidth]{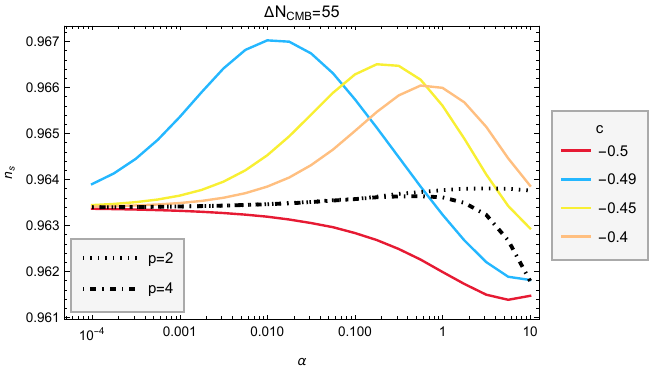}
  \caption{Scalar spectral index, $n_s$, shown as a function of $\alpha$ for four binomial models~\eqref{eq:potential with two terms} with different values of $c<0$, see legend. 
  Differently from Fig.~\ref{fig: ns binomial with c<0}, here $n_s$ is computed for fixed $\DNCMB=55$. 
  Note these numerical results are the same as those displayed in the right panel of Fig.~\ref{fig:xCMB-ns-comparison}, with $s$ defined in Eq.~\eqref{eq: s definition}. 
  The black lines correspond to results obtained for monomial T-models~\eqref{eq: potential monomial T-models} with $p=2$ (dotted) and $p=4$ (dot-dashed). }
  \label{fig: ns binomial with c<0 and DNCMB fixed}
\end{figure}

In Fig.~\ref{fig: ns binomial with c<0} we observe that the $\alpha$-dependence of $n_s$ for models with $c\simeq-1/2$ is qualitatively different with respect to that of monomial models.
This was not the case for models with $c>0$, see Fig.~\ref{fig:ns-alpha}. 
One might wonder whether this is due to the dependence of $\DNCMBinst$ on the potential parameters, see Fig.~\ref{fig: DN_CMB inst rh binomial models}. 
By computing $n_s$ for fixed $\DNCMB=55$ and four $-0.5\leq c \leq -0.4$ models we show in Fig.~\ref{fig: ns binomial with c<0 and DNCMB fixed} that this is not the case.
Moreover, one observes in Fig.~\ref{fig: ns binomial with c<0} that the variation in $n_s$ with respect to monomial predictions is larger with respect to that of $c>0$ models, being up to $\text{few}\times \mathcal{O}(10^{-3})$. 
These observations suggest that when $c\simeq -1/2$ the potential parameters $\alpha$ and $c$ enter the large-$\DNCMB$ expansion of $n_s$ as a leading-order effect, or at least earlier with respect to when the $p$ and $\alpha$ dependence appears for monomial models, see Eqs.~\eqref{eq: ns monomial} and~\eqref{eq: ns universal monomial}. 
To demonstrate this, we compute in App.~\ref{app: analytical predictions for ns with c~1/2} the analytical expression for $n_s$ in the large-$\DNCMB$ expansion for models with $c\equiv -1/2+s$ and $0\leq s\ll 1$. 
While we are not able to obtain an expression valid everywhere in the parameter space, we provide $n_s$ results at $\mathcal{O}(\DNCMB^{-2})$ and $\mathcal{O}(s^2)$, which correctly track the numerical values when $s=\{0, \, 0.01\}$ and $0.1\leq \alpha \leq 4$, see Fig.~\ref{fig:largeN-performance}. 
In the following, we will highlight few interesting results from App.~\ref{app: analytical predictions for ns with c~1/2}. 

The analytical expression for $n_s$ when $s=0$ ($c=-1/2$) is 
\begin{equation}
    n_s 
    = 
    1
    -
    \frac{2}{\DNCMB}
    -
    \frac{3^{3/2}\sqrt{\alpha}}{2^{5/2}\DNCMB^{3/2}} 
    +
     \mathcal{O}\left( \DNCMB^{-2}\right)\;,
\end{equation}
see Eq.~\eqref{eq:ns zero}. 
At leading-order in $\DNCMB$ the $n_s$ predictions are universal, i.e. insensitive to the potential parameters $\alpha$ and $c$. 
Nevertheless, the leading correction is of $\mathcal{O}(\DNCMB^{-3/2})$ and it is therefore larger than the $\mathcal{O}(\DNCMB^{-2})$ leading correction obtained for monomial models, see Eq.~\eqref{eq: ns universal monomial}. 
Moreover, it is proportional to $\sqrt{\alpha}$.
These two facts combined explain why $n_s$ displays a larger variation against changes in $\alpha$ for models with $c=-1/2$ with respect to monomial ones. 

For a model with $s=0.01$ ($c=-0.49$), representative of models with $c\to -1/2$, we provide in Eq.~\eqref{eq:ns-dominant-terms} an analytical expression for $n_s$, valid for $0.1\lesssim \alpha\lesssim4$,
\begin{equation}
  n_s 
  \approx
  1
  -
  s^2
  \,
  \frac{2.370}{\alpha} 
   +
  s \,
  \frac{1.089}{\sqrt{\alpha}\, \DNCMB^{1/2}}
  -
  \frac{2}{\DNCMB}
  -
 \frac{3^{3/2}\sqrt{\alpha}}{2^{5/2}\DNCMB^{3/2}} 
  +
  \frac{0.426+1.915\,\alpha^{0.957}}{\DNCMB^2}
  \;. 
\end{equation}
This correctly captures the rise in $n_s$ observed in Fig.~\ref{fig: ns binomial with c<0 and DNCMB fixed} when $\alpha$ decreases below 4. 
We observe that on top of the baseline $s=0$ result, which includes the aforementioned $\mathcal{O}(\DNCMB^{-3/2})$ contribution, having $s\neq 0$ induces two additional terms.   
First, a $\mathcal{O}(\DNCMB^0)$-term appears, inversely proportional to $\alpha$. 
Second, the leading-order term in the large-$\DNCMB$ expansion is now of $\mathcal{O}(\DNCMB^{-1/2})$, also inversely proportional to a power of $\alpha$. 

Overall, we conclude from App.~\ref{app: analytical predictions for ns with c~1/2} that the $n_s$ predictions for models with $c\simeq -1/2$ are qualitatively different from those of monomial models. 
This is either due to a larger sub-leading term of $\mathcal{O}(\DNCMB^{-3/2})$ for $c=-1/2$ models, or to the appearance of leading terms before the usual $\mathcal{O}(\DNCMB^{-1})$ for models with $c>-1/2$. 
For $s\neq 0$ models the $n_s$ predictions are no longer universal (i.e. independent of the potential parameters $\alpha$ and $c$) at leading-order in the large-$\DNCMB$ expansion.  

\begin{figure}
  \centering
  \includegraphics[width=0.46\textwidth]{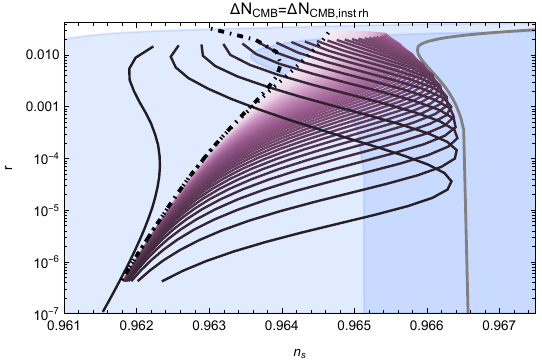}
  \includegraphics[width=0.45\textwidth]{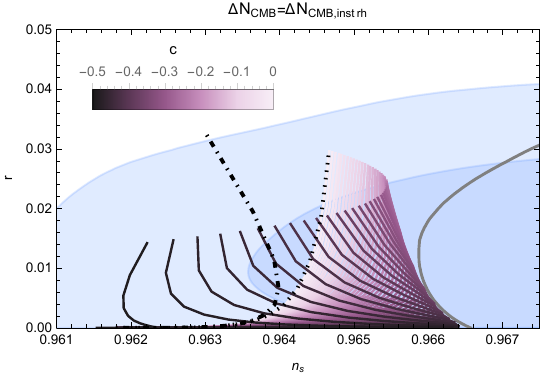}
  \caption{Numerical predictions in the $(n_s, \, r)$ plane for binomial models~\eqref{eq:potential with two terms} with $-1/2\leq c<0$. 
  Values of $n_s$ and $r$ have been computed numerically from Eqs.~\eqref{eq:nssecond} and~\eqref{eq:r second} under the assumption of instantaneous reheating, $\DNCMB=\DNCMBinst$. 
  Moving from bottom (smaller $r$) to top (larger $r$) along each fixed-$c$ coloured line corresponds to increasing $\alpha$ from $10^{-4}$ to $10$.
  The black lines show $n_s$ and $r$ values computed for monomial T-models~\eqref{eq: potential monomial T-models} with $p=2$ (dotted) and $p=4$ (dot-dashed).
  Results for $n_s$ and $r$ are shown together with the SPA+BK (blue) and SPA+BK+DESI (gray) $68\%$ and $95\%$ C.L. contours~\cite{Balkenhol:2025wms}.
  In the left (right) panel $r$ is shown on a logarithmic (linear) scale.}
  \label{fig: ns and r binomial with c<0}
\end{figure}

In Fig.~\ref{fig: ns and r binomial with c<0} we display predictions for models with $-1/2\leq c<0$ and $10^{-4}\leq \alpha\leq 10$ in the $(n_s,\,r)$ plane, computed by assuming instantaneous reheating. 
The tensor-to-scalar ratio is mainly determined by the value of $\alpha$. 
Nevertheless, for fixed $\alpha$ the variation for different $c$ values is larger than for models with $c>0$, see Fig.~\ref{fig:ns-r binomial models with c>0}. 
For example, for two models with $\alpha=10$ and $c=-0.01$ and $c=-1/2$ we obtain $r=0.29$ and $r=0.014$ respectively, with an approximate $50\%$ relative variation of the smaller one if compared against the larger one. 

Differently from models with $c>0$, models with $-1/2<c<0$ can lead to $n_s$ values larger than those of monomial potentials with $p=2$. 
This can be seen by comparing Fig.~\ref{fig: ns binomial with c<0} against Fig.~\ref{fig:ns-alpha}. 
So while in Fig.~\ref{fig:ns-r binomial models with c>0} the $(n_s,\,r)$ predictions were always on the \textit{left} of the monomial $p=2$ ones, we now have predictions for the majority of the models considered being on the \textit{right} of the $p=2$ line. 
For this reason, a large number of binomial models is now compatible with SPA+BK at $68\%$ C.L.. 
Moreover, the predictions now follow closely the SPA+BK+DESI $95\%$ C.L. contour. 
Nevertheless, they do not enter it, since as $c$ approaches $-1/2$ the value of $\alpha$ at which $n_s$ is the largest decreases, yielding a correspondingly smaller and smaller $r$.

\section{Reheating and its equation of state}
\label{sec: reheating}

In this section we study the post-inflationary dynamics of the binomial potential~\eqref{eq:potential with two terms}.
Our purpose is to determine the impact of $c$ on the equation of state parameter during the first few e-folds of perturbative reheating. 
 
After the end of inflation, the homogeneous inflaton oscillates around the minimum of its potential and it decays into Standard Model particles, leaving behind a radiation dominated universe. 
In this work we assume that the reheating process can be described perturbatively, i.e. the inflaton is weakly coupled to other fields. 
By summarizing the effect of all decay channels in a total decay rate $\Gamma_{\rm tot}$, the homogeneous field obeys
\begin{equation}
\label{eq: inflaton eom reheating}
  \ddot\phi+\left(3H+\Gamma_{\rm tot}\right)\dot\phi+V_{\phi}=0 \;. 
\end{equation}
At the beginning of reheating one typically has $H\gg\Gamma_{\rm tot}$, and only when $H$ drops sufficiently below $\Gamma_{\rm tot}$ particle production becomes efficient.  
Due to the weak coupling between the inflaton and other particles, it might takes many e-folds before this condition is realised. 
As will be made clear below, any effect from $c$ is confined to the first few e-folds of reheating. 
For this reason, we will assume $H\gg \Gamma_{\rm tot}$ in the following, simplify Eq.~\eqref{eq: inflaton eom reheating} accordingly, and assume that the inflaton dominates the total energy density and pressure, i.e. $P_\text{tot}\approx P_\phi, \, \rho_\text{tot}\approx \rho_\phi$. 

As done in Eq.~\eqref{eq:DNCMB-general}, the reheating stage can be parametrised in terms of its duration, $\DNrh \equiv N_\text{rh} - N_\text{end}$, and its average equation of state (EoS) parameter 
\begin{equation}
\label{eq: eos reheating}
  \bar w
  \equiv
  \frac{1}{\DNrh}
  \int_{N_{\rm end}}^{N_{\rm rh}}\!\mathrm{d}N'\,w(N') \;, 
\end{equation}
where $N_{\rm rh}$ marks the end of reheating and $w(N)\equiv {P_\text{tot}(N)}/{\rho_\text{tot}(N)}$ is the instantaneous EoS parameter.   
Because no explicit decay model is considered, $\DNrh$ is treated as a free parameter.  
Nevertheless, as said above we will focus on the first few e-folds of reheating. 
In this regime it is useful to introduce the average EoS parameter evaluated at an intermediate time $N_{\rm end}<N<N_\text{rh}$,
\begin{equation}
\label{eq:wcumulative}
  \bar w(N)
  \equiv
  \frac{1}{N-N_{\rm end}}
  \int_{N_{\rm end}}^{N}\mathrm{d}N'\,w(N') \;. 
\end{equation}
Note that the over-bar symbol distinguishes averaged EoS parameters from the instantaneous value $w(N)$.

\begin{figure}
  \centering
  \includegraphics[width=0.55\textwidth]{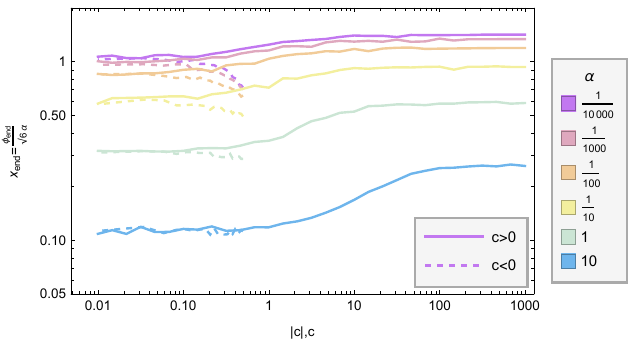}
  \caption{Field displacement at the end of inflation, $\xend=\phi_\text{end}/\sqrt{6\alpha}$, as a function of $c\geq-1/2$ for binomial potentials~\eqref{eq:potential with two terms}. 
  Solid (dashed) curves correspond to $c>0$ ($c<0$), and each one of the coloured lines correspond to a different value of $\alpha$, see the colour legend.}
  \label{fig: xend reheating}
\end{figure}

The EoS generated by coherent inflaton oscillations depends on the shape of the potential explored by the field. 
We mark the end of inflation when $\epsilon_H$ reaches unity, and denote the corresponding value of the inflaton as $\xend\equiv \phi_\text{end}/\sqrt{6\alpha}$.
We display in Fig.~\ref{fig: xend reheating} the numerical value of $\xend$ computed for models with $c\geq-1/2$ and $10^{-4}\leq\alpha\leq10$. 
For fixed $\alpha$, the dependence on $c$ is mild. 
On the other hand, for fixed $c$ the dependence on $\alpha$ is appreciable; increasing $\alpha$ reduces $x_{\rm end}$, while it flattens as $\alpha$ is lowered from $10^{-1}$ to $10^{-4}$.  
Motivated by these trends, from now on we focus on the three values $\alpha=\{0.1, \, 1,\,10\}$. 

After the end of inflation, $|x|\equiv |\phi|/\sqrt{6\alpha}\leq \xend \lesssim1$. 
Therefore we can expand Eq.~\eqref{eq:potential with two terms} for $|x|<1$, yielding
\begin{equation}
\label{eq: potential at small field}
  \frac{V(x)}{V_0}
  =
  x^2
  +
  \left(c-\frac{2}{3}\right)x^4+\mathcal{O}(x^6) \;. 
\end{equation}
A large value of $x_{\rm end}$ and/or a large coefficient $c$ might lead to the quartic term dominating initially over the quadratic one.  
Nevertheless, the field value is time-dependent, with a decreasing oscillation amplitude. 
Therefore, even when the quartic correction dominates initially, the quadratic term must eventually take over as $|x|\rightarrow0$.  
When the inflaton oscillations occur over a monomial potential, $V(x)\propto|x|^p$, averaging over many oscillations gives the effective EoS parameter~\cite{Turner:1983he}
\begin{equation}
    \label{eq:monomial w}
  \bar w=\frac{p-2}{p+2} \;. 
\end{equation}
For example, this is the case for monomial T-models~\eqref{eq: potential monomial T-models}.
A quadratic minimum ($p=2$) yields matter domination $(\bar w =0)$, while a quartic one ($p=4$) to radiation domination $(\bar w=1/3)$. 
The binomial potential is therefore expected to pass from a radiation-like stage (if possible at all) to a matter-like stage as the amplitude of the inflaton oscillations decreases.  
Note that this description is only asymptotic: at intermediate times both terms in Eq.~\eqref{eq: potential at small field} contribute and the potential cannot be represented by a monomial power.

In Sec.~\ref{sec:evolvingw} we numerically compute $\bar w(N)$ for three models with $c=500$ and $\alpha=\{0.1,1,10\}$, thereby explicitly showing the time-dependence of the EoS parameter as outlined above. 
We then perform a systematic parameter-space exploration to establish how large $c$ needs to be in order to realise a transient reheating stage with $\bar w\geq 1/3$ in Sec.~\ref{sec: fine tuning for w>=1/3}.

\subsection[\texorpdfstring{A time-dependent $\bar w$}{A time-dependent w}]{A time-dependent $\bm{\bar w}$}
\label{sec:evolvingw}

\begin{figure}
  \centering
  \includegraphics[width=0.45\textwidth]{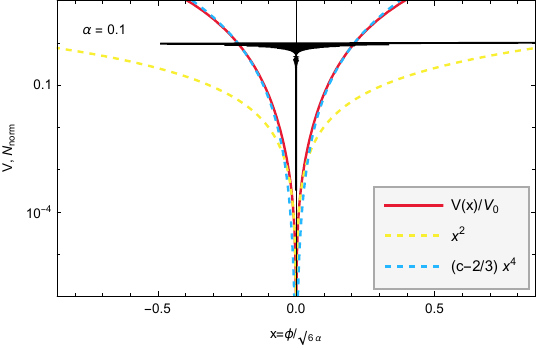}
  \centering
  \includegraphics[width=0.45\textwidth]{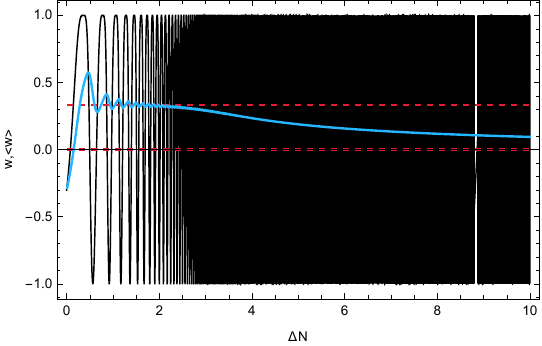}
  \centering
  \includegraphics[width=0.45\textwidth]{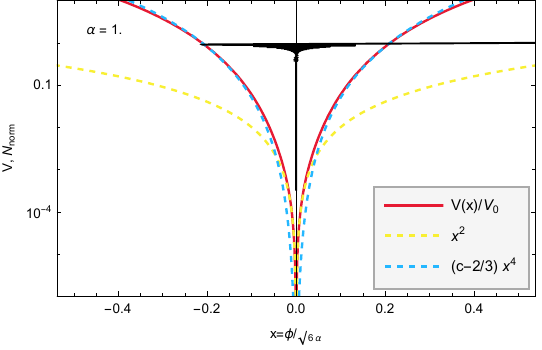}
  \centering
  \includegraphics[width=0.45\textwidth]{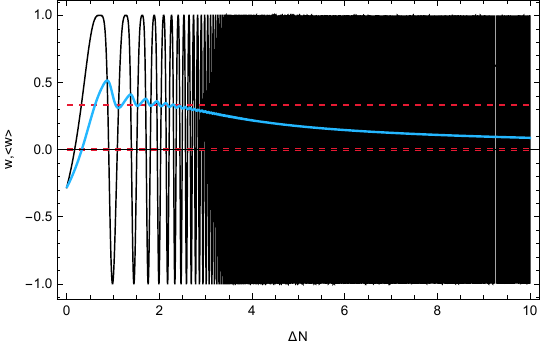}
  \centering
  \includegraphics[width=0.45\textwidth]{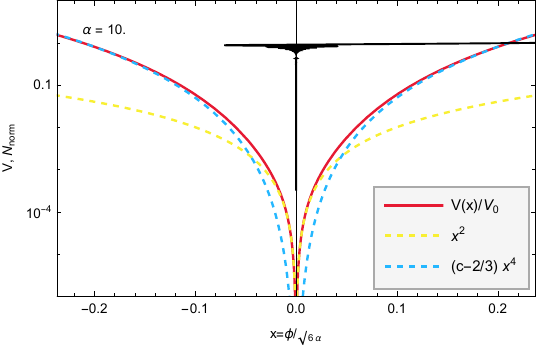}
  \centering
  \includegraphics[width=0.45\textwidth]{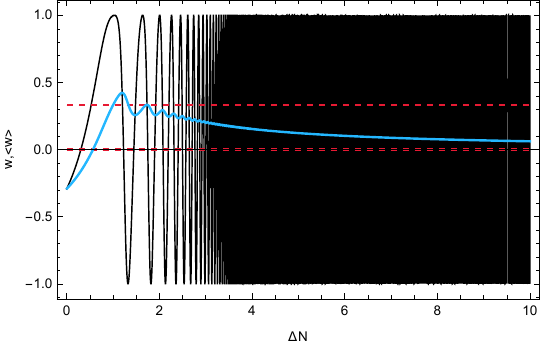}
  \caption{Post-inflationary evolution of the inflaton and EoS parameter computed numerically for three models with $c=500$ and $\alpha=\{0.1, \, 1,\, 10\}$.  
  Each row corresponds to a different value of $\alpha$. 
  In the left panels we display the evolution of $x\equiv \phi/\sqrt{6\alpha}$ for the first 10 e-folds after inflation, shown against a suitably normalised e-folding variable, $N_\text{norm}$. 
  In particular, $N_{\rm norm}=1$ at the end of inflation and then decreases towards zero as time goes by. 
  The inflaton trajectory is superimposed on the dimensionless potential $V(x)/V_0$ (red) and its quadratic (yellow, dashed) and quartic (blue, dashed) contributions.  
  In the right panels we show the instantaneous EoS parameter $w(N)$ (black) and the averaged one $\bar w(N)$ (blue), see Eq.~\eqref{eq:wcumulative}. 
  The red, dashed lines highlight $w=0$ and $w=1/3$.}
  \label{fig: w time-dependence}
\end{figure}

For three models with a sizeable amplitude of the quartic term, $c=500$, and $\alpha=\{0.1,1,10\}$ we numerically solve the background equations of motion, see Eq.~\eqref{eq: inflaton eom reheating} with $H\gg \Gamma_\text{tot}$, for 10 e-folds after the end of inflation.  
The inflaton values at the end of inflation are respectively $x_\text{end} =\{0.961,\, 0.598,\, 0.264\}$. 
Numerical results are displayed in Fig.~\ref{fig: w time-dependence}. 
For $\alpha=0.1$, in which case $x_\text{end}$ is the largest, the field initially oscillates with a larger amplitude, thereby exploring a region of the potential which is well approximated by the quartic contribution.
As a consequence, $\bar w(N)$ initially oscillates around $\bar w = 1/3$. 
As time goes by, the oscillation amplitude decreases and the inflaton first explores a region which is well approximated by the superposition of the quadratic and quartic terms, and finally enters the quadratic portion, causing $\bar w(N)$ to decrease gradually towards $\bar w =0$. 
The same sequence of events also occurs for $\alpha=1$, but in this case the quartic-dominated interval is shorter because $x_{\rm end}$ is smaller.  
For $\alpha=10$, the oscillation amplitude at the end of inflation is substantially reduced, and the system only briefly displays $\bar w (N) \sim 1/3$, without really settling into this regime. 
Note that the transition between radiation-like to matter-like reheating is smooth in all three cases, due to the potential effectively containing both the quadratic and quartic contributions throughout the crossover region.

\subsection[\texorpdfstring{A sustained $\bar w\geq 1/3$ stage}{A sustained w>= 1/3 stage}]{A sustained $\bm{\bar w\geq 1/3}$ stage}
\label{sec: fine tuning for w>=1/3}

We have established in Sec.~\ref{sec:evolvingw} that depending on the values of $c$ and $\alpha$ there might be an initial phase with $\bar w\sim 1/3$ before the quadratic term takes over and $\bar w \to 0$. 
How large $c \equiv c_4/c_2$ must be for the quartic term to dominate the EoS parameter for an appreciable number of e-folds during reheating?
To answer this question we systematically scan models with $\alpha=\{10^{-2},10^{-1},1,10\}$ and $10\leq c\leq10^5$.
Note that negative values of $c$ are not included because the potential-monotonicity condition $c\geq -1/2$ does not allow for an appreciable hierarchy between the quartic and quadratic terms when $c<0$. 
Each model is evolved for up to five e-folds after the end of inflation, and $\bar w(N)$ is computed over this range. 
As in Fig.~\ref{fig: w time-dependence}, $\bar w(N)$ is highly oscillatory. 
In order to robustly assess the number of e-folds for which $\bar w(N) \geq 1/3$ we first smooth the $\bar w(N)$ data. 
This is achieved by passing the data through a low-pass-filter with cut-off frequency $1/2 \, \omega_\text{max}$, where $\omega_\text{max}$ is the frequency associated with the Fourier coefficient with largest amplitude. 

\begin{figure}
  \centering
  \includegraphics[width=0.55\textwidth]{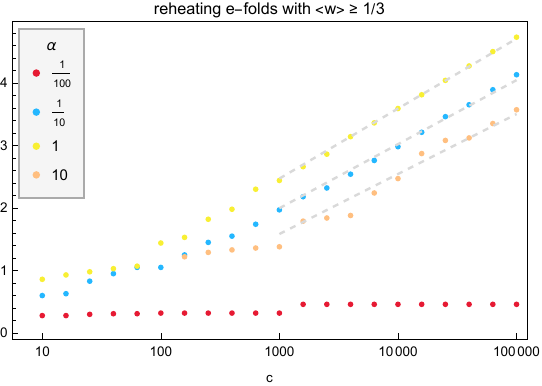}
  \caption{Number of post-inflationary e-folds for which the smoothed $\bar w(N)$, see Eq.~\eqref{eq:wcumulative}, satisfies $\bar w \geq1/3$, shown as a function of $c$.  
  Different colours correspond to different $\alpha$ values, see the point legend. 
  The dashed, gray lines are obtained by fitting the underlying numerical data in the large-$c$ regime with linear functions of $\ln c$.
  If the smoothed $\bar w$ never crossed the $1/3$ threshold, there is no coloured point displayed for the corresponding model.}
  \label{fig: fine tuning c}
\end{figure}

Results for the duration of the interval over which the smoothed $\bar w(N)$ remains at or above $1/3$, $\Delta N_{\bar w\geq 1/3}$, are displayed in Fig.~\ref{fig: fine tuning c}. 
For $\alpha=10^{-2}$, $\Delta N_{\bar w\geq 1/3}$ remains a fraction of an e-fold throughout the whole range of $c$.  
For $\alpha\geq0.1$, $\Delta N_{\bar w\geq 1/3}$ grows with $c$, with an initial scattered trend (due to the oscillatory nature of $\bar w(N)$) before settling into a linear function of $\ln c$ at at large $c$. 
Interestingly, the slope of the data in this regime depends on $\alpha$ only mildly; by fitting the numerical data with the function $\Delta N_{\bar w\geq 1/3} = a + b \ln c$, we obtain $b\simeq0.42-0.49$. 

From Fig.~\ref{fig: fine tuning c} we observe that obtaining a quartic-dominated reheating stage lasting around four e-folds requires $c\equiv {c_4}/{c_2}\sim 10^5$ for $\alpha \gtrsim0.1$.  
For the supergravity potential~\eqref{eq: potential full series} this constitutes a substantial coefficient hierarchy.  
Moreover, because the quadratic term always dominates sufficiently close to the origin when $c_2\neq0$, the radiation-like stage is necessarily transient.

\section{Discussion}
\label{sec:conclusions}
In this work we consider $\alpha$-attractor T-models beyond the usual monomial formulation~\eqref{eq: potential monomial T-models}. 
By working with the binomial potential in Eq.~\eqref{eq:potential with two terms} we investigate what effects might arise from the interaction between quadratic and quartic terms during inflation and reheating. 
We assume $c\geq -1/2$, which ensures monotonicity of $V(\phi)$ and avoids potential problems with the initial conditions.

We recover the large-scale predictions of monomial potentials with $p=2$ and $p=4$ when $c\to 0$ and $c\to \infty$ respectively. 
For $c>0$ we find that $n_s$ and $r$ values interpolate non-monotonically between the two asymptotic regimes. 
Nevertheless, the deviation from monomial predictions for $n_s$ is at most of $\mathcal{O}(10^{-3})$ and the $\alpha$-dependence is qualitatively unchanged. 
We explain these numerical findings by showing that at leading order in the large-$\DNCMB$ expansion the $n_s$ predictions are universal, see Eq.~\eqref{eq: universal predictions} and Eq.~\eqref{eq: ns large DNCMB c>0}. 

On the other hand, for models with $c\simeq -1/2$ we find a larger variation of $n_s$ with respect to the monomial results, as well as a different $\alpha$-dependence.
In other words, the parameters $c$ and $\alpha$ affect the $n_s$ predictions much more than in the case $c>0$, as confirmed by the large-$\DNCMB$ expansion in Eq.~\eqref{eq:ns-dominant-terms}. 
When $0<s\equiv c+1/2 \ll 1$ two new terms appear at $\mathcal{O}(\DNCMB^0)$ and $\mathcal{O}(\DNCMB^{-1/2})$, which depend on $s$ and (inversely) on powers of $\alpha$. 
At leading order in the large-$\DNCMB$ expansion the predictions are therefore no longer universal. 
When $c=-1/2$ ($s=0$) we find that the deviation with respect to the monomial behavior is due to the leading-order correction to Eq.~\eqref{eq: universal predictions} being of $\mathcal{O}(\DNCMB^{-3/2})$, instead of $\mathcal{O}(\DNCMB^{-2})$ as usual. 

We explore the interplay between the quadratic and quartic terms in Eq.~\eqref{eq:potential with two terms} during the first few e-folds of perturbative reheating. 
If $c$ is large enough and/or the inflaton's oscillations start off with a large enough amplitude, the quartic term initially dominates the dynamics before the quadratic one eventually takes over as the oscillations amplitude decreases. 
We show that the binomial potential~\eqref{eq:potential with two terms} might therefore lead to a time-dependent EoS during reheating, with $\bar w\sim 1/3$ initially before $\bar w \to 0$. 
Nevertheless, even a few e-folds of radiation-like reheating require a substantial hierarchy between the quartic and quadratic terms, e.g. we obtain 4 e-folds with $\bar w\geq 1/3$ if $c\sim 10^5$ and $\alpha\gtrsim0.1$. 

Earlier works exist on the interplay between different potential terms during reheating and time-dependence of $\bar w$. 
In Ref.~\cite{Turner:1983he} Turner considers the perturbative effect of higher-order powers in a polynomial potential.
In Ref.~\cite{Saha:2020bis} Saha \textit{et al.} demonstrate for potentials which behave like $V(\phi)\propto \phi^p$ near the minimum that growing inhomogeneties and interactions with a lighter scalar field yield a time-dependent EoS approaching 1/3 after a few e-folds when $p\geq 3$.
In Ref.~\cite{Clery:2024dlk} Clery \textit{et al.} point out that a bare mass term added to a quartic potential would lead to $\bar w$ transitioning from 1/3 to 0.
We provide here the first numerical study of a time-dependent $\bar w$ due to the interplay between a quadratic and a quartic term, where the latter is potentially much larger than the former. 
Furthermore, this effect is discussed in the context of recent debates on the consistency of monomial T-models with SPA+BK+DESI measurements, see below.  

The binomial model we consider represents the first instance of a more generic potential derived within supergravity, see Eq.~\eqref{eq: potential full series}.
Our results show that by working with a monomial T-model~\eqref{eq: potential monomial T-models} with $p=4$ one might miss non-trivial behavior of the large-scale observables. 
Furthermore, assuming that $\bar w$ is uniquely determined by the quartic term for an extended period relies on the assumption of substantial tuning of the underlying supergravity potential~\eqref{eq: potential full series}. 
For example, by extrapolating the linear fits in Fig.~\ref{fig: fine tuning c}, one finds that 10 e-folds of perturbative reheating with $\bar w \geq 1/3$ require $c=10^9-10^{11}$. 

For monomial T-models, $p=6$ is the smallest value of $p$ leading to stiff reheating $(\bar w=1/2)$, and able to reconcile these potentials with current SPA+BK+DESI data if $\DNrh\gtrsim 24$, see Figs.~1,~2 and~5 in Ref.~\cite{Iacconi:2025odq}.
For this reason, a logical next step would be to investigate the trinomial potential
\footnote{See Refs.~\cite{Dalianis:2018frf,Iacconi:2021ltm} for earlier studies of T-model polynomial potentials up to sixth order. Differently from Eq.~\eqref{eq:potential with three terms}, they include also odd powers of the hyperbolic tangent, and were introduced to implement an inflection point ($V_{\phi\phi}(\phi_\text{infl})=0$), which can lead to non-attractor behavior and consequent enhancement of the power spectrum on short scales.}
\begin{equation}
\label{eq:potential with three terms}
  V(\phi)
  = c_2
  \tanh^2\left(\frac{\phi}{\sqrt{6\alpha}}\right)
  + c_4\,\tanh^4\left(\frac{\phi}{\sqrt{6\alpha}}\right)
  + c_6\,\tanh^6\left(\frac{\phi}{\sqrt{6\alpha}}\right) \;. 
\end{equation}
Does such a long reheating stage dominated by the sixth power require a large hierachy of parameters? 
We expect our argument on the time-dependence of $\bar w$ to apply to the potential in Eq.~\eqref{eq:potential with three terms} as well, as it relies on the inevitable decrease of the inflaton's oscillations amplitude. 
Therefore we anticipate that only a very large hierarchy of the coefficients might decrease the tension with current large-scale data.  
Nevertheless, as was the case for binomial models~\eqref{eq:potential with two terms} with $c\simeq -1/2$, there could be areas of the parameter space $(\alpha, \, c_4,\, c_6 )$ which lead to non-universal behavior for $n_s$, potentially lifting it towards unity without the need for an extended phase of stiff reheating. 
We leave this exploration for future work. 

\section*{Acknowledgments}
The author would like to thank David Mulryne and David Wands for careful reading of the manuscript and very useful comments. 
This work was supported by the Science and Technology Facilities Council (grant number ST/X000931/1).
For the purpose of open access, the author has applied a Creative Commons Attribution (CC-BY) licence to any Author Accepted Manuscript version arising from this work.
Supporting research data are available on reasonable request. 
The author used generative AI tools to support proofreading of the manuscript. 

\appendix
\section[\texorpdfstring{Analytical ${\bm{n_s}}$ and its large-$\bm{\DNCMB}$ expansion}{Analytical ns and its large-DN_CMB expansion}]{Analytical ${\bm{n_s}}$ and its large-$\bm{\DNCMB}$ expansion}
\label{app: analytical predictions for ns}

In this Appendix we derive analytical predictions for $n_s$ in $\alpha$-attractor T-model binomial potentials, see Eq.~\eqref{eq:potential with two terms}. 
Our aim is to establish at which order in the large-$\DNCMB$ expansion the new potential parameter $c$ enters the predictions for $n_s$. 
We consider the case of $c>0$ in App.~\ref{app: analytical predictions for ns with c>0}, and $c\simeq  -1/2$ in App.~\ref{app: analytical predictions for ns with c~1/2}. 

We first derive an analytical expression for $\phi_\text{end}\equiv \phi(N_\text{end})$, obtained by solving 
\begin{equation}
\label{eq: phi_end condition}
    \epsilon_V(\phi_\text{end})=1 \;. 
\end{equation}
Due to the break-down of slow-roll towards the end of inflation, this criterion is not good enough for computing precise predictions~\cite{Auclair:2024udj}. 
Nevertheless, in the main body of this work we do not rely on analytical computations for $n_s$, and rather use numerical solutions. 
For our purpose in this Appendix the criterion~\eqref{eq: phi_end condition} suffices. 

By inverting the slow-roll equation of motion for the inflaton
\begin{equation}
    \label{eq: phi_CMB condition}
    \DNCMB \equiv N_\text{end}-N_\text{CMB} \simeq \int_{\phi_\text{end}}^{\phiCMB} \mathrm{d}\phi \, \frac{V}{V_\phi}  \;, 
\end{equation}
we then compute $\phiCMB\equiv \phi(N_\text{CMB})$.

Finally, we substitute $\phiCMB$ in the (first-order) slow-roll expression for $n_s$, and obtain an analytical prediction for $n_s$. 
To check the analytical expressions obtained for $\phi_\text{end},\,\phiCMB$ and $n_s$ we compare them against their numerical counterparts. 
We then give the series expansion of $n_s$ for large $\DNCMB$.
The final results are quoted in Eq.~\eqref{eq: ns large DNCMB c>0} for $c>0$ and in Eqs.~\eqref{eq: ns zero explicit alpha dependence},~\eqref{eq:ns first in s} and~\eqref{eq:ns second in s} for $c\simeq -1/2$. 
While our computational strategy is standard (see e.g. Appendix A of Ref.~\cite{Iacconi:2021ltm} for monomial T-models), the more complex form of the potential requires additional approximations and novel approaches to make the different tasks analytically tractable.  

\subsection[\texorpdfstring{Case $c>0$}{Case c>0}]{Case $\bm{c>0}$}
\label{app: analytical predictions for ns with c>0}

\begin{figure}
  \centering
  \includegraphics[width=0.5\textwidth]{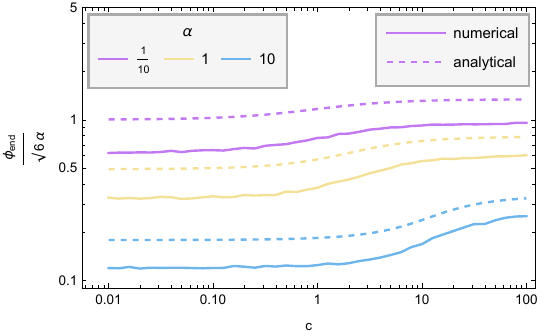}
  \caption{Comparison between the numerical value of $\phi_{\rm end}/\sqrt{6\alpha}$ (continuous line) and the analytical estimate~\eqref{eq: phi_end c>0} obtained from Eq.~\eqref{eq: phi_end condition} (dashed line). }
  \label{fig:phiend}
\end{figure}

By using Eq.~\eqref{eq: phi_end condition}, we obtain 
\begin{equation}
\label{eq: phi_end c>0}
  \phi_{\rm end}
  = \sqrt{\frac{3\alpha}{2}}\,\ln \left[ z(c,\,\alpha) \right] \;, 
\end{equation}
where $z(c,\alpha)$ is the $4\textit{th}$-root of $p(z;\, c,\,\alpha)=0$ with 
\begin{multline}
\label{eq:rootpoly}
    p(z; \,c,\, \alpha)=3\alpha+6c\alpha+3c^2\alpha
 +(12\alpha-12c^2\alpha)z 
+(-16-64c-64c^2+12\alpha-24c\alpha+12c^2\alpha)z^2
\\
+(-64+256c^2-12\alpha+12c^2\alpha)z^3
+(-96+128c-384c^2-30\alpha+36c\alpha-30c^2\alpha)z^4
\nonumber\\
+(-64+256c^2-12\alpha+12c^2\alpha)z^5
+(-16-64c-64c^2+12\alpha-24c\alpha+12c^2\alpha)z^6
\nonumber\\
+(12\alpha-12c^2\alpha)z^7
 +(3\alpha+6c\alpha+3c^2\alpha)z^8 \;. 
\end{multline}
We compare the analytical expression in Eq.~\eqref{eq: phi_end c>0} against numerical results in Fig.~\ref{fig:phiend}. 
Eq.~\eqref{eq: phi_end c>0} correctly reproduces the qualitative $c$-dependence of the numerical results. 
We do not expect the agreement to be exact, due to slow-roll breaking down towards the end of inflation, see the discussion below Eq.~\eqref{eq: phi_end condition}.  
Note that Eq.~\eqref{eq: phi_end c>0} is derived by assuming $c>0$, and cannot therefore be used to recover the known $c=0$ result, corresponding to a T-model with $p=2$. 

\begin{figure}
  \centering
  \includegraphics[width=0.48\textwidth]{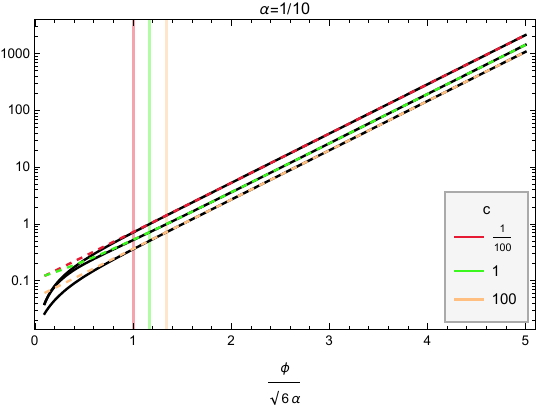}
  \includegraphics[width=0.48\textwidth]{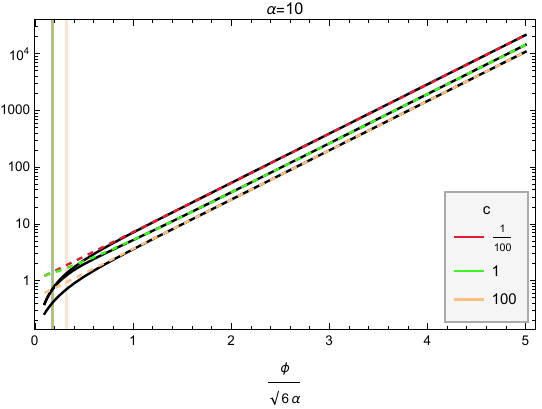}
  \caption{Comparison between the full integrand in Eq.~\eqref{eq: phi_CMB condition}, $V/V_{\phi}$, (black, continuous line), and the large-field approximation in Eq.~\eqref{eq:integrandseries} (colored, dashed line). 
  The case of $\alpha=1/10$ $(\alpha=10)$ is displayed in the left (right panel), and in each panel colored lines correspond to different values of $c$. 
  For each model considered, vertical lines denote $x_{\rm end}$, and during inflation $x > x_{\rm end}$.}
  \label{fig:integrandseries}
\end{figure}

To compute $\phiCMB$, one needs to perform the integral in Eq.~\eqref{eq: phi_CMB condition} and then analytically invert the resulting expression.
Due to the complexity of the potential, the latter is not possible. 
To circumvent this issue, we observe that typically during inflation $x\equiv \phi/\sqrt{6\alpha}\gg1$. 
We therefore expand the integrand in Eq.~\eqref{eq: phi_CMB condition} for large $y\equiv \exp{x}$, and truncate the series at next-to-leading order 
\begin{equation}
  \frac{V}{V_{\phi}}
  =
  \frac{\sqrt{3\alpha /2}\,(1+c)}{4(1+2c)}y^2
  +
  \frac{\sqrt{3\alpha/2}\,c}{(1+2c)^2} + \mathcal{O}\left(y^{-2}\right) \;.  
  \label{eq:integrandseries}
\end{equation}
The full integrand is compared against Eq.~\eqref{eq:integrandseries} in Fig.~\ref{fig:integrandseries} for some representative values of $\alpha$ and $c$.  
One sees that the approximation works well over most of the parameter space, but becomes less accurate for large $\alpha$ close to $x_{\rm end} \equiv \phi_{\rm end}/\sqrt{6\alpha}$.
This is because the larger $\alpha$ the smaller $\xend$ is, see Fig.~\ref{fig:phiend}, thereby putting the corresponding models outside of the region of validity of Eq.~\eqref{eq:integrandseries}. 

\begin{figure}
  \centering
  \includegraphics[width=0.43\textwidth]{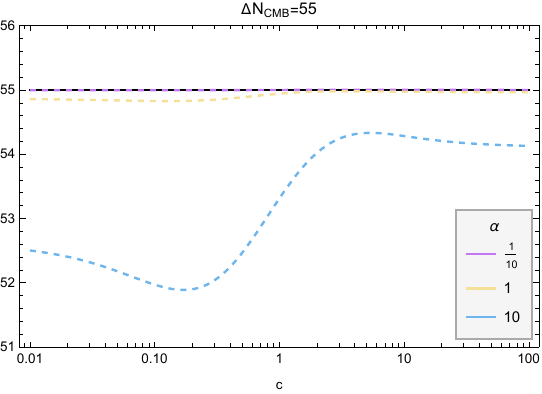}
  \includegraphics[width=0.48\textwidth]{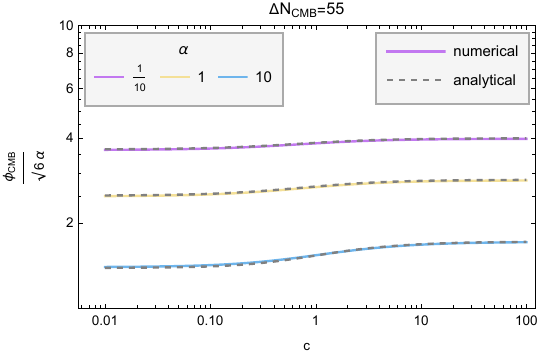}
  \caption{
  \textit{Left panel:} 
  Consistency test of the solution in Eq.~\eqref{eq: phi_CMB}. 
  The approximate solution~\eqref{eq: phi_CMB} is inserted into the exact e-fold integral, Eq.~\eqref{eq: phi_CMB condition}, and compared against $\Delta N_{\rm CMB}=55$. 
  Deviations for large $\alpha$ reflect the reduced accuracy of the large-field integrand approximation~\eqref{eq:integrandseries} near $\phi_{\rm end}$.
  \textit{Right panel:}
  Comparison between numerical (continuous, coloured line) and analytical (dashed, grey line) predictions for $\phi_{\rm CMB}/\sqrt{6\alpha}$. 
  }
  \label{fig: phi_CMB plots}
\end{figure}

By substituting Eq.~\eqref{eq:integrandseries} in Eq.~\eqref{eq: phi_CMB condition} and massaging the resulting expression, one obtains 
\begin{equation}
\label{eq: Lambert 1}
    A \exp{\left(2 \xCMB\right)} + B\,  \left( 2 \xCMB \right) = C \;,  
\end{equation}
where $\xCMB\equiv \phiCMB/\sqrt{6\alpha}$ and 
\begin{subequations}
\begin{align}
  A &= \frac{3\alpha(1+c)}{8(1+2c)} \;, 
  \label{eq:adef}\\
  B &= \frac{3c\alpha}{2(1+2c)^2} \;, 
  \label{eq:bdef}\\
  C &= \Delta N_{\rm CMB}
  + \frac{3\alpha (1+c) (1+2c) \exp{\left(2\xend\right)} + 24 c \alpha \xend}{8 (1+2c)^2} \;. 
  \label{eq:cdef}
\end{align}
\end{subequations}
Eq.~\eqref{eq: Lambert 1} can be solved in terms of the Lambert function, $W_k(z)$.
For real numbers and for $\alpha>0$ and $c>0$, the solution is given in terms of the first branch only, $W_0(z)$,
\begin{equation}
\label{eq: phi_CMB}
  \phi_{\rm CMB}
  = \sqrt{\frac{3\alpha}{2}}
  \left\{
  \frac{C}{B}
  - W_0\!\left[\frac{A}{B}\exp\left(\frac{C}{B}\right)\right]
  \right\} \;. 
\end{equation}
In the left panel of Fig.~\ref{fig: phi_CMB plots} we test the solution~\eqref{eq: phi_CMB} by substituting it in Eq.~\eqref{eq: phi_CMB condition}, where the full integrand has been retained. 
For illustration, we set $\DNCMB=55$. 
While it was obtained by means of the approximation~\eqref{eq:integrandseries}, Eq.~\eqref{eq: phi_CMB} provides a good solution for the full integral when $\alpha\lesssim 1$. 
On the other hand, when $\alpha=10$ its performance worsens and it includes a spurious dependence on $c$. 
This is due to the fact that towards the end of inflation Eq.~\eqref{eq:integrandseries} no longer correctly represents the full integrand $V/V_\phi$ for these models, see Fig.~\ref{fig:integrandseries}. 

In the right panel of Fig.~\ref{fig: phi_CMB plots} the analytical solution~\eqref{eq: phi_CMB} is compared against numerical results.
The agreement is excellent, with the relative difference between the two remaining within $1.1\%$ over the parameter space tested.

\begin{figure}
    \centering
  \includegraphics[width=0.48\textwidth]{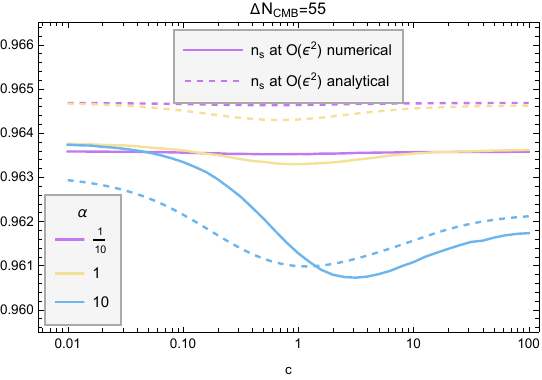}
  \includegraphics[width=0.48\textwidth]{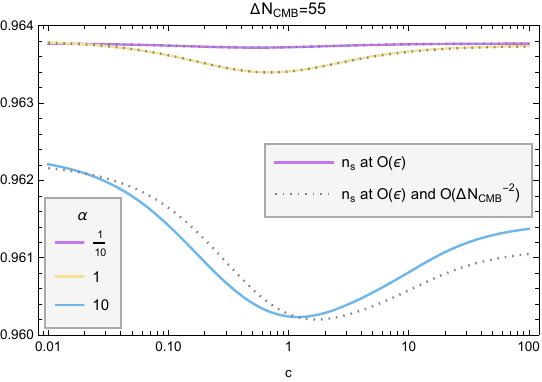}
  \caption{
  \textit{Left panel:} 
  Comparison between numerical results for $n_s$ at second-order in slow-roll and the analytical predictions obtained by using Eqs.~\eqref{eq: phi_end c>0} and~\eqref{eq: phi_CMB}.
  \textit{Right panel:}
  Comparison between the first-order slow-roll analytical prediction for $n_s$ and its expansion at $\mathcal{O}\left( {\Delta N_{\rm CMB}}^{-2}\right)$, Eq.~\eqref{eq: ns large DNCMB c>0}.}
  \label{fig: ns analytical c>0}
\end{figure}

By substituting Eqs.~\eqref{eq: phi_end c>0} and~\eqref{eq: phi_CMB} in Eq.~\eqref{eq:nssecond}, one obtains the analytical expression for $n_s$ at second-order in the slow-roll expansion. 
The expression is lengthy and not particularly illuminating, and we therefore do not include it here. 
In the left panel of Fig.~\ref{fig: ns analytical c>0} we compare it against the corresponding numerical results. 
The analytical expression reproduces the qualitative dependence on $c$, with a better performance for $\alpha\lesssim 1$.  
The residual vertical shift relative to the numerical values can be attributed to the combined effect of employing slow-roll approximations up to the end of inflation, Eqs.~\eqref{eq: phi_end condition} and~\eqref{eq: phi_CMB condition}, and using the large-field approximation Eq.~\eqref{eq:integrandseries} to determine $\phi_{\rm CMB}$.

Since we are interested in the large-$\Delta N_{\rm CMB}$ expansion of $n_s$, we can work at leading order in slow roll, see the $\mathcal{O}(\epsilon)$ terms in Eq.~\eqref{eq:nssecond}. 
Due to the presence of $W_0(z)$ in Eq.~\eqref{eq: phi_CMB}, expanding directly $n_s$ is cumbersome. 
To circumvent this problem, we expand $n_s$ for large $\exp{\left( 2 \xCMB\right)}$ (with $\phiCMB$ left implicit), and separately $\exp{\left( 2 \xCMB\right)}$ for large $\DNCMB$. 
By substituting the latter into the former, and expanding again for large $\DNCMB$, we obtain 
\begin{multline}
\label{eq: ns large DNCMB c>0}
  n_s = 1-\frac{2}{\Delta N_{\rm CMB}}
  +\frac{1}{{\Delta N_{\rm CMB}}^2 4(1+2c)^2}
  \Bigg[
  3 \alpha (1+c) (1+2c)
  \exp{\left(2 \xend\right)}
  +24\alpha c\,x_{\rm end}
    \\
  -6\alpha\left(1+c(4c+\ln 64)\right)
  +12\alpha c \ln\left(
  \frac{3\alpha(1+c)}{(1+2c)\Delta N_{\rm CMB}}
  \right)
  \Bigg]
  +\mathcal{O}\left({\Delta N_{\rm CMB}}^{-3}\right) \;. 
\end{multline}
At leading order we recover the $\alpha$-attractor universal predictions, and model-dependent corrections first appear at $\mathcal{O}\left({\Delta N_{\rm CMB}}^{-2}\right)$. 
In the right panel of Fig.~\ref{fig: ns analytical c>0} we compare the full analytical expression for $n_s$ at first-order in slow-roll against Eq.~\eqref{eq: ns large DNCMB c>0}, showing that the latter correctly includes the qualitative dependence on the potential parameters $\alpha$ and $c$. 
The result in Eq.~\eqref{eq: ns large DNCMB c>0} explains why the predictions of the binomial model~\eqref{eq:potential with two terms} with $c>0$ do not dramatically deviate from those of $p=2$ and $p=4$ T-models: at leading order in $\DNCMB$ they are universal. 

Note that Eq.~\eqref{eq: ns large DNCMB c>0} has been obtained by means of the series expansion in Eq.~\eqref{eq:integrandseries}. 
As displayed in Fig.~\ref{fig:integrandseries}, the performance of such approximation varies across the parameter space. 
For this reason, one should not use Eq.~\eqref{eq: ns large DNCMB c>0} to obtain precise predictions for $n_s$. 

\subsection[\texorpdfstring{Case $c\simeq-1/2$}{Case c~-1/2}]{Case $\bm{c\simeq-1/2}$}
\label{app: analytical predictions for ns with c~1/2}

\begin{figure}
  \centering
  \includegraphics[width=0.48\textwidth]{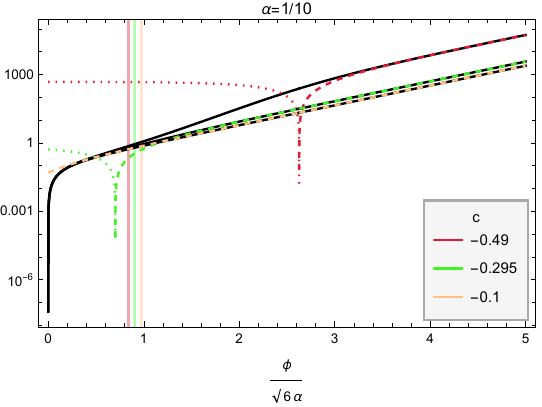}
  \includegraphics[width=0.48\textwidth]{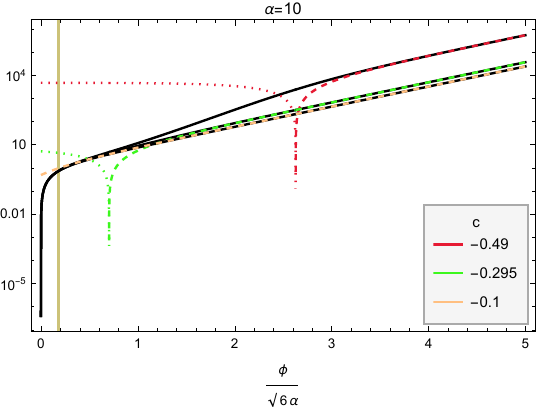}
  \caption{Comparison between the integrand in Eq.~\eqref{eq: phi_CMB condition}, $V/V_{\phi}$, (black, continuous line) and its large-field approximation~\eqref{eq:integrandseries} (colored, dashed line) for models with $\alpha=1/10$ (left panel) and $\alpha=10$ (right panel), and different values of $-1/2<c<0$, as detailed in the legend.   
  Vertical lines mark $x_{\rm end}$, computed for each model by using Eq.~\eqref{eq: phi_end condition}.}
  \label{fig: series integrand}
\end{figure}

Before focussing on $c\simeq -1/2$, let us comment on generic models with $-1/2\leq c<0$. 
To obtain analytical predictions for $n_s$, one could proceed as done in App.~\ref{app: analytical predictions for ns with c>0} for $c>0$, by computing $\phi_\text{end}$ from Eq.~\eqref{eq: phi_end condition}, and $\phi_\text{CMB}$ from Eq.~\eqref{eq: phi_CMB condition} by means of the large-field expansion of the integrand $V/V_\phi$. 
Note that in Eq.~\eqref{eq:integrandseries} the factor $(1+2c)$ appears at the denominator, indicating that the case $c=-1/2$ must be treated separately. 
In Fig.~\ref{fig: series integrand} we compare $V/V_\phi$ against its large-field expansion for models with $\alpha=\{1/10,10\}$ and three values of $-1/2<c<0$. 
The integrand $V/V_\phi$ can be reliably approximated during inflation ($x\geq x_\text{end}$) only when $|c|\ll1$.
The phenomenology of models with $|c|\ll1$ is of limited interest (see Sec.~\ref{sec: binomial potential during inflation}), and we therefore do not pursue their analytical treatment further, and move on to the case $c\simeq -1/2$.   

Since one cannot rely on the large-field expansion of $V/V_\phi$, see Eq.~\eqref{eq:integrandseries}, we propose an alternative approach to  treat models with $c\simeq -1/2$. 
By defining the parameter
\begin{equation}
    \label{eq: s definition}
    s \equiv c + 1/2 >0 \;, 
\end{equation}
one could instead perform series expansions for small $s$. 
For example, the inflaton value at the end of inflation can be defined as 
\begin{equation}
  \frac{\phi_\text{end}}{\sqrt{6\alpha}} \equiv x_{\rm end}=x_{\rm end}^{(0)}+s\,x_{\rm end}^{(1)}+s^2x_{\rm end}^{(2)}+\mathcal{O}(s^3) \;, 
  \label{eq:xend-series}
\end{equation}
where the $k$-\textit{th} term is found by solving Eq.~\eqref{eq: phi_end condition} at $\mathcal{O}(s^k)$. 
By expanding Eq.~\eqref{eq: phi_end condition} in $s$ up to second order one obtains schematically 
\begin{equation}
  f\left(x_{\rm end}^{(0)},\alpha\right)
  +
  s 
  \,
  g\left(x_{\rm end}^{(0)},x_{\rm end}^{(1)},\alpha\right)
  +
  s^2
  \, 
  h\left(x_{\rm end}^{(0)},x_{\rm end}^{(1)},x_{\rm end}^{(2)},\alpha\right) + \mathcal{O}(s^3)
  =
  1 \;, 
  \label{eq:xend-hierarchy}
\end{equation}
and $x_{\rm end}^{(0)}$, $x_{\rm end}^{(1)}$, $x_{\rm end}^{(2)}$ are obtained by solving 
\begin{equation}
    f\left(x_{\rm end}^{(0)},\alpha\right) = 1\;, \quad g\left(x_{\rm end}^{(0)},x_{\rm end}^{(1)},\alpha\right)= h\left(x_{\rm end}^{(0)},x_{\rm end}^{(1)},x_{\rm end}^{(2)},\alpha\right)=0 \;.
\end{equation}
Note that the $k$-\textit{th} order solution depends on all the preceding ones. 
The resulting analytical expressions for $x_{\rm end}^{(0)}$, $x_{\rm end}^{(1)}$, $x_{\rm end}^{(2)}$ are not particularly illuminating, and we therefore do not include them here. 

\begin{figure}
  \centering
  \includegraphics[width=0.5\textwidth]{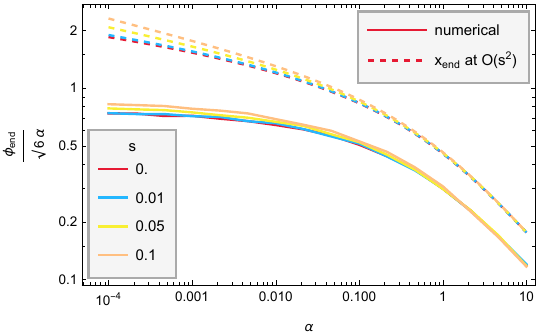}
  \caption{Comparison between numerical values of $x_{\rm end}$ (continuous line) and the analytical solution in Eq.~\eqref{eq:xend-series} (dashed line), obtained at $\mathcal{O}(s^2)$.
  Results are obtained for four different $s$-models, see the colour legend, and shown against $\alpha$.}
  \label{fig:xend-series}
\end{figure}

In Fig.~\ref{fig:xend-series} we compare Eq.~\eqref{eq:xend-series} against numerical results for models with different $s$. 
The qualitative dependence on $\alpha$ and $s$ is correctly reproduced. 
Nevertheless, we observe a vertical shift between numerical and analytical results.
The inaccuracy of the latter is due to the break-down of slow-roll towards the end of inflation, and not to the series-expansion procedure, as it is the case also for models with $s=0$.
While we keep all terms up to $\mathcal{O}(s^2)$ in Eq.~\eqref{eq:xend-series}, we have checked that the second- and third-order terms do not significantly contribute. 

In analogy with $x_{\rm end}$, we expand $x_{\rm CMB}$ as
\begin{equation}
  \frac{\phi_{\rm CMB}}{\sqrt{6\alpha}} \equiv
  x_{\rm CMB}
  =
  x_{\rm CMB}^{(0)}
  + 
  s
  \,
  x_{\rm CMB}^{(1)}
  +
  s^2
  \,
  x_{\rm CMB}^{(2)}
  +
  \mathcal{O}(s^3) \;.
  \label{eq:xCMB-series}
\end{equation}
Each term is computed by expanding Eq.~\eqref{eq: phi_CMB condition} in $s$ up to second-order,
\begin{multline}
    m\left( 
    x_{\rm CMB}^{(0)},\, x_{\rm end}^{(0)},\, \alpha
    \right) 
    + 
    s
    \,
    n\left( 
    x_{\rm CMB}^{(0)},\, x_{\rm CMB}^{(1)},\, x_{\rm end}^{(0)},\, x_{\rm end}^{(1)},\, \alpha
    \right)
    +
    \\
    s^2
    \, 
    o\left(
    x_{\rm CMB}^{(0)},\, x_{\rm CMB}^{(1)},\, x_{\rm CMB}^{(2)},\, x_{\rm end}^{(0)},\, x_{\rm end}^{(1)},\, x_{\rm end}^{(2)},\, \alpha
    \right) 
    +
    \mathcal{O}(s^3)
    = 
    \DNCMB \;, 
\end{multline}
and by solving 
\begin{equation}
    \begin{split}
    m\left( 
    x_{\rm CMB}^{(0)},\, x_{\rm end}^{(0)},\, \alpha
    \right) &= \DNCMB
    \;, 
    \\
    n\left( 
    x_{\rm CMB}^{(0)},\, x_{\rm CMB}^{(1)},\, x_{\rm end}^{(0)},\, x_{\rm end}^{(1)},\, \alpha
    \right) &= 0\;, \\
    o\left(
    x_{\rm CMB}^{(0)},\, x_{\rm CMB}^{(1)},\, x_{\rm CMB}^{(2)},\, x_{\rm end}^{(0)},\, x_{\rm end}^{(1)},\, x_{\rm end}^{(2)},\, \alpha
    \right) &=0 \;.
\end{split}
\end{equation}
Note that the solutions for $x_{\rm CMB}^{(0)}$, $x_{\rm CMB}^{(1)}$ and $x_{\rm CMB}^{(2)}$ also depend on those in Eq.~\eqref{eq:xend-series}. 

\begin{figure}
  \centering
  \includegraphics[width=0.44\textwidth]{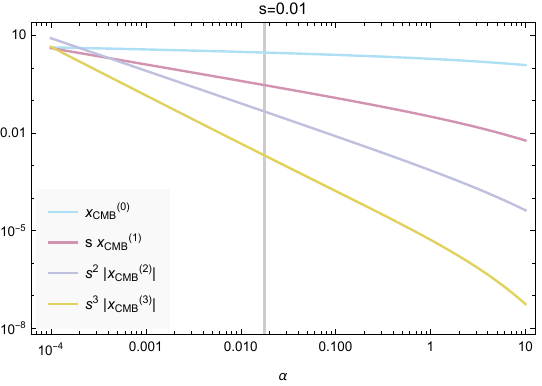}
  \includegraphics[width=0.44\textwidth]{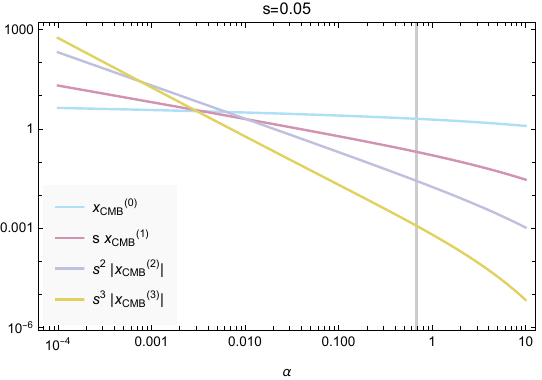}
  \includegraphics[width=0.44\textwidth]{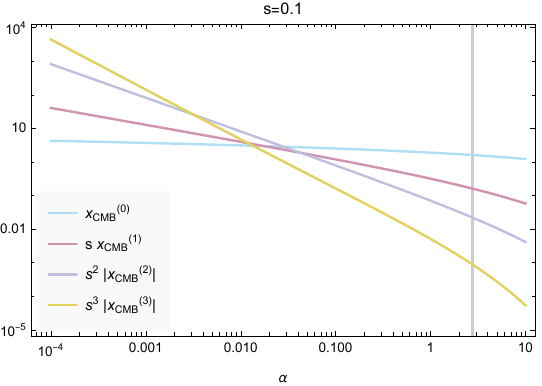}
  \caption{Zeroth-, first-, second- and third-order solutions for $x_{\rm CMB}$, see Eq.~\eqref{eq:xCMB-series}, shown against $\alpha$.  
  Each panel corresponds to a different $s$-model. 
  The gray, vertical line marks the convergence threshold $\alpha_{\rm conv}$, see the main text for its definition.}
  \label{fig:xCMB-convergence}
\end{figure}

As for $x_{\rm end}$, we do not include the resulting expressions here, but display each term in the series expansion~\eqref{eq:xCMB-series} in Fig.~\ref{fig:xCMB-convergence} for three different $s$-models.  
Here one can compare the magnitude of each successive contribution to $x_{\rm CMB}$.
Notably, the series expansion does not converge for arbitrarily small $\alpha$. 
We define the minimum $\alpha$ for which the series~\eqref{eq:xCMB-series} is convergent, $\alpha_{\rm conv}$, as the value above which the first-order contribution, $s \, x_{\rm CMB}^{(1)}$, drops below $0.1 \times x_{\rm CMB}^{(0)}$. 
The smaller $s$ is, the larger the convergence region of the series expansion. 
Even if the criterion defining $\alpha_{\rm conv}$ does not technically ensure convergence of the series, Fig.~\ref{fig:xCMB-convergence} shows that the series is indeed convergent for $\alpha\gtrsim\alpha_{\rm conv}$. 
In Fig.~\ref{fig:xCMB-convergence} we also show the third-order component of $x_{\rm CMB}$, even if this is not explicitly included in Eq.~\eqref{eq:xCMB-series}. 
By comparing second- and third-order contributions, one sees that when the series is convergent, it can be truncated at $\mathcal{O}(s^2)$. 

\begin{figure}
    \centering
    \includegraphics[width=0.44\textwidth]{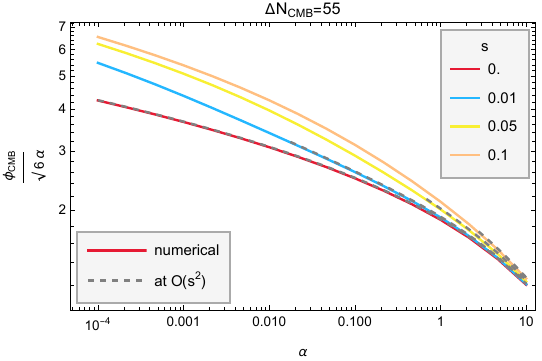}
    \centering
    \includegraphics[width=0.44\textwidth]{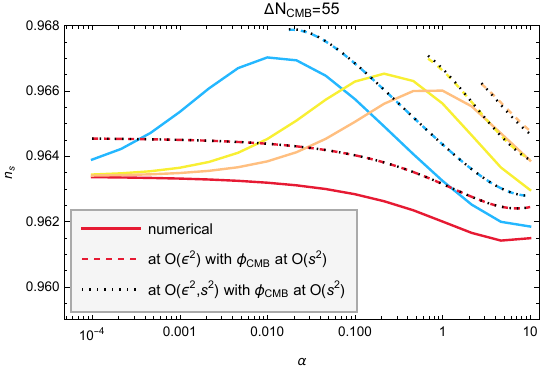}
    \caption{\textit{Left panel:} Comparison between analytical solutions for $x_{\rm CMB}$ at $\mathcal{O}(s^2)$, see Eq.~\eqref{eq:xCMB-series}, (gray, dashed line) and numerical results (coloured line).
    These are displayed for four different $s$-models, see the colour legend. 
    For each $s$, analytical results are shown for $\alpha\gtrsim \alpha_{\rm conv}$, ensuring thar the series is convergent, see Fig.~\ref{fig:xCMB-convergence}.  
    \textit{Right panel:} 
    Comparison between numerical results for $n_s$ at second-order in the slow-roll expansion (coloured line), analytical results computed by using Eq.~\eqref{eq:xCMB-series} (coloured, dashed line), and analytical results where before substituting Eq.~\eqref{eq:xCMB-series} $n_s$ is expanded at $\mathcal{O}(s^2)$ (gray, dotted line). 
    See the left panel for the colour legend identifying the different $s$ models.}
    \label{fig:xCMB-ns-comparison}
\end{figure}

The second-order analytical expression for $x_{\rm CMB}$~\eqref{eq:xCMB-series} is compared against numerical results in the left panel of Fig.~\ref{fig:xCMB-ns-comparison}.  
For illustration, we have fixed $\DNCMB=55$. 
For $\alpha\gtrsim\alpha_{\rm conv}$ the analytical results agree with the numerical ones to within $0.7\%$ throughout the range of models considered. 

Analytical results for $n_s$ at second-order in the slow-roll expansion are obtained by substituting Eqs.~\eqref{eq:xCMB-series} and~\eqref{eq:xend-series} in Eq.~\eqref{eq:nssecond}.
Results are displayed in the right panel of Fig.~\ref{fig:xCMB-ns-comparison}, where for each $s$-model the analytical expressions are only shown for $\alpha\geq \alpha_{\rm conv}$, i.e. when the series~\eqref{eq:xCMB-series} converges. 
By comparing numerical and analytical results, one observes that the latter correctly reproduce the qualitative dependence of the former on $\alpha$ and $s$. 
Nevertheless, even if the $x_{\rm CMB}$ analytical results are within $0.7\%$ of their numerical counterparts, these small differences cause the analytical predictions for $n_s$ to systematically shift upwards with respect to the numerical results. 
For consistency with the series-expansion approach employed to obtain $x_{\rm end}$ and $x_{\rm CMB}$, we have also included $n_s$ obtained by first expanding Eq.~\eqref{eq:nssecond} up to $\mathcal{O}(s^2)$ and then substituting Eq.~\eqref{eq:xend-series} and~\eqref{eq:xCMB-series}. 
These are basically undistinguishable from the lines obtained without expanding in $s$.  

While our series-expansion approach does not provide analytical results for $n_s$ over the whole parameter space --specifically only for $\alpha\geq \alpha_{\rm conv}$--, they capture the rise in $n_s$ observed for each $s$-model for decreasing values of $\alpha$. 
For the purpose of understanding the origin of the $n_s$ increase, one can therefore expand for large $\DNCMB$ the analytical predictions we obtained.
It suffices to use the expression for $n_s$ at first-order in slow-roll as we have checked that its $\alpha$-dependence is qualitatively the same as that of the second-order results in Fig.~\ref{fig:xCMB-ns-comparison}. 
Since the $\alpha$ values over which the series~\eqref{eq:xCMB-series} converges are more restricted the larger $s$ is, we decide to focus on the cases $s=0$ and $s=0.01$.  

By substituting Eqs.~\eqref{eq:xend-series} and~\eqref{eq:xCMB-series} in the $\mathcal{O}(\epsilon)$ expression for $n_s$, see Eq.~\eqref{eq:nssecond}, and expanding up to second order in $s$, we schematically obtain 
\begin{equation}
\label{eq:ns-series}
    n_s= n_s^{(0)} + s\, n_s^{(1)} + s^2 \, n_s^{(2)} + \mathcal{O}(s^3) \;.  
\end{equation}
We do not include expressions for $n_s^{(0)}$, $n_s^{(1)}$ and $n_s^{(2)}$ as they are very long and not particularly useful at this stage. 
For each term in Eq.~\eqref{eq:ns-series} we perform a large-$\DNCMB$ expansion up to $\mathcal{O}(\DNCMB^{-2})$. 

\begin{figure}
    \centering
    \includegraphics[width=0.5\textwidth]{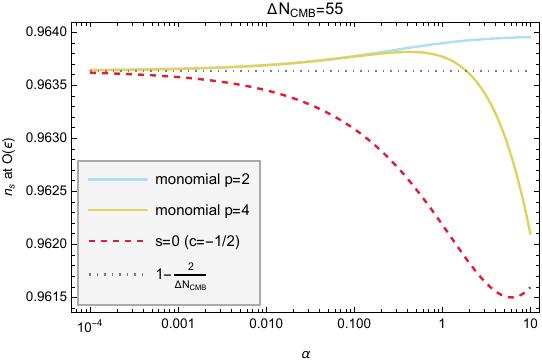}
    \caption{Scalar spectral index, $n_s$, at first-order in the slow-roll expansion displayed against $\alpha$ for two monomial T-models (continuous, coloured lines), see Eq.~\eqref{eq: ns monomial}, and for the binomial model~\eqref{eq:potential with two terms} with $s=0$ $(c=-1/2)$ (dashed, red line). 
    These results have been computed for fixed $\DNCMB=55$. 
    A dotted, grey line marks the universal $\alpha$-attractor predictions.}
    \label{fig:ns for s=0 against monomial models}
\end{figure}

The zeroth-order contribution, corresponding to models with $s=0$ $(c=-1/2)$, is 
\begin{equation}
    \label{eq:ns zero}
    n_s^{(0)} = 
    1
    -
    \frac{2}{\DNCMB}
    -
    \frac{3^{3/2}\sqrt{\alpha}}{2^{5/2}\DNCMB^{3/2}} 
    +
    \frac{3 \alpha \left[5 + 12 \cosh(2 x_{\rm end}^{(0)}) + \cosh(4 x_{\rm end}^{(0)})\right]}
     {32\,\DNCMB^2} 
     +
     \mathcal{O}\left( \DNCMB^{-5/2}\right)\;. 
\end{equation}
Since the binomial potential under scrutiny, Eq.~\eqref{eq:potential with two terms}, is obtained as the superposition of two monomial T-models~\eqref{eq: potential monomial T-models} with $p=2$ and $p=4$, it is useful to compare this expression against the one obtained at first-order in slow-roll for these monomial potentials
\begin{equation}
\label{eq: ns monomial}
    n_s \simeq 1
    -
    \frac{
    2 \DNCMB
    +
    \frac{\sqrt{3\alpha\left(3\alpha+p^2\right)}}{p} +\frac{3\alpha}{2}
    }{
    \DNCMB^2
    +
    \frac{\DNCMB\sqrt{3\alpha\left(3\alpha+p^2\right)}}{p}
    +\frac{3\alpha}{4}
    } \;, 
\end{equation} 
which for large $\DNCMB$ is 
\begin{equation}
\label{eq: ns universal monomial}
    n_s
    \simeq  
    1
    -
    \frac{2}{\DNCMB}
    +
    \frac{-\frac{3}{2}\alpha +\frac{\sqrt{3\alpha\left(p^2+3\alpha\right)}}{p}}
{\DNCMB^2}
    +
    \mathcal{O}\left( \DNCMB^{-3}\right)\;.
\end{equation}
First, in Eq.~\eqref{eq:ns zero} we observe that at leading order we recover the universal predictions. 
Nevertheless, having $s=0$ $(c=-1/2)$ modifies the leading correction with respect to the one obtained for monomial T-models.  
In Eq.~\eqref{eq:ns zero} this is of $\mathcal{O}(\DNCMB^{-3/2})$, and therefore larger than the $\mathcal{O}(\DNCMB^{-2})$ correction in Eq.~\eqref{eq: ns universal monomial}. 
Moreover, over the range of $\alpha$ considered, the coefficient of the leading correction is always negative in Eq.~\eqref{eq:ns zero}, while in Eq.~\eqref{eq: ns universal monomial} it is always positive for $p=2$ and for $p=4$ turns from positive to negative for increasing $\alpha$. 
Both in Eq.~\eqref{eq:ns zero} and Eq.~\eqref{eq: ns universal monomial} the leading correction is measured by $\alpha$, and we expect to recover the universal predictions for $\alpha\to0$. 
Fig.~\ref{fig:ns for s=0 against monomial models} illustrates the effects discussed above. 
In order to isolate them from the potential-dependence of $\DNCMB$~\cite{Iacconi:2023mnw}, we fix $\DNCMB=55$. 
The larger leading correction in Eq.~\eqref{eq:ns zero} and the fact that it contributes with a negative sign explain why in Fig.~\ref{fig:ns for s=0 against monomial models} we observe that for models with $s=0$ $n_s$ is always smaller than that of monomial potentials with $p=2$ and $p=4$.
Note that the numerical results in Fig.~\ref{fig: ns binomial with c<0} display a more complex dependence on $\alpha$ with respect to those in Fig.~\ref{fig:ns for s=0 against monomial models}. 
This is due to the fact that they also include the dependence of $\DNCMB$ on the potential parameters. 

In Eq.~\eqref{eq:ns zero}, the $\alpha$-dependence of the $\mathcal{O}(\DNCMB^{-2})$ term is implicit, as $x_{\rm end}^{(0)}$ itself depends on $\alpha$. 
To specify it, we must resort to numerical evaluation of $x_{\rm end}^{(0)}$ for a range of $\alpha$ values. 
As for $n_s^{(0)}$, also for higher-order contributions to Eq.~\eqref{eq:ns-series} we will have to pursue a mixture of analytical and numerical evaluation. 
We decide to work in the reduced parameter range $0.1\leq \alpha \leq 10$, over which Eq.~\eqref{eq:xCMB-series} is always convergent for $s=0.01$, see Fig.~\ref{fig:xCMB-convergence}.  
We evaluate the coefficient of $\DNCMB^{-2}$ in Eq.~\eqref{eq:ns zero} for 40 values of $0.1\leq \alpha \leq 10$, and fit the results. 
Substituting the fit in Eq.~\eqref{eq:ns zero} makes the $\alpha$-dependence explicit
\begin{equation}
\label{eq: ns zero explicit alpha dependence}
  n_s^{(0)}
  =
  1
  -
  \frac{2}{\DNCMB}
  -
 \frac{3^{3/2}\sqrt{\alpha}}{2^{5/2}\DNCMB^{3/2}} 
  +
  \frac{0.426+1.915\,\alpha^{0.957}}{\DNCMB^2}
  +
  \mathcal{O}\left( \DNCMB^{-5/2}\right)\;. 
\end{equation}

Beyond zeroth-order, the $n_s$ analytical expressions are so complicated that it is not possible to perform directly the large $\DNCMB$ expansion. 
As anticipated, we adopt instead a semi-numerical strategy.
First, we evaluate $n_s^{(1)}$ and $n_s^{(2)}$ for 40 values of $0.1\leq \alpha \leq 10$ and series-expand up to $\mathcal{O}(\DNCMB^{-2})$ the resulting expressions. 
For example, for $n_s^{(1)}$ this yields schematically 
\begin{equation}
    n_s^{(1)} = 
    \frac{\mathfrak{f}}{\DNCMB^{1/2}}
    + 
    \frac{\mathfrak{g}}{\DNCMB}
    + 
    \frac{\mathfrak{h}}{\DNCMB^{3/2}}
    +
    \frac{\mathfrak{i}}{\DNCMB^2}
    +
    \mathcal{O}(\DNCMB^{-5/2}) \;, 
\end{equation}
where $\mathfrak{f, \, g, \, \ldots}$ are numerical coefficients which depend on $\alpha$. 
The numerical data is then fitted to return analytical expressions $\mathfrak{f}(\alpha), \, \mathfrak{g}(\alpha), \, \ldots$. 
This procedure yields 
\begin{equation}
\label{eq:ns first in s}
    n_s^{(1)} 
    = 
    \frac{1.089}{\sqrt{\alpha}\, \DNCMB^{1/2}}
    +
    \frac{-0.344-0.765\,\alpha^{0.538}}{\DNCMB^{3/2}}
    +
    \frac{0.245-1.695\,\alpha^{0.987}}{\DNCMB^2}
    +
    \mathcal{O}(\DNCMB^{-5/2})
\end{equation}
and 
\begin{equation}
\label{eq:ns second in s}
    n_s^{(2)} 
    = 
    -\frac{2.370}{\alpha}
    +
    \frac{1.905}{\sqrt{\alpha}\,\DNCMB^{1/2}}
    +
    \frac{-0.833-0.907\,\alpha^{0.596}}{\DNCMB^{3/2}}
    +
    \frac{0.014+2.873\,\alpha}{\DNCMB^2}
    +
    \mathcal{O}(\DNCMB^{-5/2})\;. 
\end{equation}
In both $n_s^{(1)}$ and $n_s^{(2)}$, the numerical coefficients of the $\mathcal{O}(\DNCMB^{-1})$ term are $\mathcal{O}(10^{-16}-10^{-14})$, and we have therefore neglected their contribution. 
Note that the $\alpha$-dependence of terms with the same $\DNCMB$ power in Eqs.~\eqref{eq:ns first in s} and~\eqref{eq:ns second in s} is similar.

By comparing Eqs.\eqref{eq:ns first in s} and~\eqref{eq:ns second in s} against Eq.~\eqref{eq: ns zero explicit alpha dependence} one can gauge the effects that having $s \neq 0$ introduces. 
First, we obtain terms proportional to $\DNCMB^{-1/2}$, which are absent in Eq.~\eqref{eq: ns zero explicit alpha dependence}. 
Note that in the large-$\DNCMB$ expansion of Eq.~\eqref{eq:ns-series} they provide the leading-order correction to constant terms, being larger than the $\mathcal{O}(\DNCMB^{-1})$ contribution in Eq.~\eqref{eq: ns zero explicit alpha dependence}. 
Second, Eq.~\eqref{eq:ns second in s} also includes a constant contribution, inversely proportional to $\alpha$.

\begin{figure}
  \centering
  \includegraphics[width=0.5\textwidth]{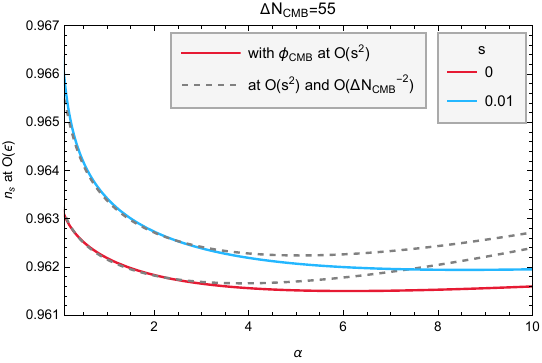}
  \caption{For models with $s=0$ and $s=0.01$, comparison between the $n_s$ results at $\mathcal{O}(s^2)$ (coloured lines) and the corresponding results obtained by further expanding $n_s$ for large $\DNCMB$ (dashed, grey lines), see Eqs.~\eqref{eq: ns zero explicit alpha dependence},~\eqref{eq:ns first in s} and~\eqref{eq:ns second in s}.  
  Note that the coloured lines are not the same as in the right panel of Fig.~\ref{fig:xCMB-ns-comparison}, as here we have used the slow-roll expression at first-order. 
  The $\alpha$ range is restricted to the values used to obtain the coefficients in Eqs.~\eqref{eq: ns zero explicit alpha dependence}, ~\eqref{eq:ns first in s} and~\eqref{eq:ns second in s}, $0.1\leq \alpha\leq 10$.}
  \label{fig:largeN-performance}
\end{figure}

\begin{figure}
  \centering
  \includegraphics[width=0.5\textwidth]{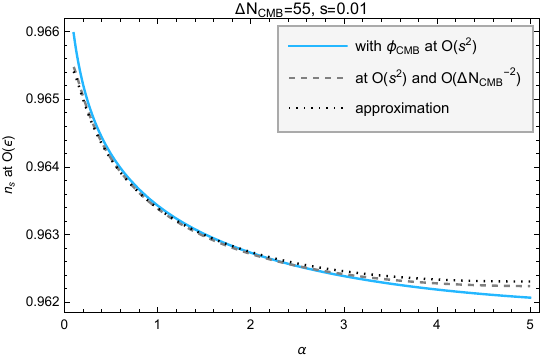}
  \caption{For models with $s=0.01$, comparison between the $n_s$ results at $\mathcal{O}(s^2)$ (light-blue line), the corresponding results obtained by further expanding $n_s$ for large $\DNCMB$ (dashed, grey line) and the approximation in Eq.~\eqref{eq:ns-dominant-terms} (dotted, gray line).}
  \label{fig:dominant-terms}
\end{figure}

In Fig.~\ref{fig:largeN-performance} we compare the analytical results for $n_s$ obtained at $\mathcal{O}(\DNCMB^{-2})$ by substituting Eqs.~\eqref{eq: ns zero explicit alpha dependence},~\eqref{eq:ns first in s} and~\eqref{eq:ns second in s} in Eq.~\eqref{eq:ns-series} against those produced without expanding for large $\DNCMB$. 
For $\alpha \gtrsim 4$ the expressions at $\mathcal{O}(\DNCMB^{-2})$ deviate significantly from the un-expanded $n_s$. 
To understand the origin of this discrepancy, we compute at each order in $s$ the percent relative error between the full expression and the one truncated at $\mathcal{O}(\DNCMB^{-2})$.
For $\alpha\lesssim1$ the relative error is below $\sim0.1\%$ for all terms. 
For larger $\alpha$, we find that the largest error is due to the $\mathcal{O}(s)$ contribution.  
For $\alpha\gtrsim5$, the relative error for the $\mathcal{O}(s)$ term can exceed the percent level, reaching roughly $10\%$ near $\alpha=10$.  
For the $\mathcal{O}(s^0)$ and $\mathcal{O}(s^2)$ terms the maximum errors are approximately $0.1\%$ and $1\%$, respectively, both near $\alpha=10$.
On the other hand, the results in Fig.~\ref{fig:largeN-performance} show that the performance of the $\mathcal{O}(\DNCMB^{-2})$ expression is very good for $\alpha\lesssim4$, and they can therefore be employed to understand the origin of the rise in $n_s$ for decreasing $\alpha$.  
For the representative case $s=0.01$ we isolate the leading contributions in Eq.~\eqref{eq:ns-series} leading to the rise in $n_s$ for decreasing $\alpha$ values, yielding 
\begin{equation}
\begin{split}
  n_s 
  &\approx
  n_s^{(0)}
  +
  s \,
  \frac{1.089}{\sqrt{\alpha}\, \DNCMB^{1/2}}
  -
  s^2
  \,
  \frac{2.370}{\alpha} \\
  &=
  1
  -
  s^2
  \,
  \frac{2.370}{\alpha} 
   +
  s \,
  \frac{1.089}{\sqrt{\alpha}\, \DNCMB^{1/2}}
  -
  \frac{2}{\DNCMB}
  -
 \frac{3^{3/2}\sqrt{\alpha}}{2^{5/2}\DNCMB^{3/2}} 
  +
  \frac{0.426+1.915\,\alpha^{0.957}}{\DNCMB^2}
  \;,
  \label{eq:ns-dominant-terms}
\end{split}
\end{equation}
where in the second line we have ordered the expression for increasing inverse-powers of $\DNCMB$. 
Fig.~\ref{fig:dominant-terms} shows that the approximation Eq.~\eqref{eq:ns-dominant-terms} correctly describes the $n_s$ increase. 
Eq.~\eqref{eq:ns-dominant-terms} demonstrates that having $s\neq 0$ induces a dependence of $n_s$ on $\alpha$ and $s$ at leading order in the large-$\DNCMB$ expansion. 
In other words, the predictions for $n_s$ are no longer universal, i.e. independent on the potential parameters.

\bibliography{refs} 
\bibliographystyle{JHEP}

\end{document}